\documentclass[11pt, a4paper]{article}

\usepackage{ifthen}

\newcommand{\MyUniPat}{lsdfgkhjvrkjlhmisdlcjn}

\makeatletter

\newcommand{\NewcommandThreeArgsTwoOpt}[5]
{
\DeclareRobustCommand#1{\@ifnextchar[%
{\csname\expandafter\@gobble\string#1@presq\endcsname}%
{\csname\expandafter\@gobble\string#1@nopresq\endcsname}}
\expandafter\def\csname\expandafter\@gobble\string#1@nopresq\endcsname##1{\@ifnextchar[%
{\csname\expandafter\@gobble\string#1@nopresq@postsq\endcsname[]{##1}}%
{\csname\expandafter\@gobble\string#1@nopresq@nopostsq\endcsname[]{##1}}}
\expandafter\def\csname\expandafter\@gobble\string#1@presq\endcsname[##1]##2{\@ifnextchar[%
{\csname\expandafter\@gobble\string#1@presq@postsq\endcsname[{##1}]{##2}}%
{\csname\expandafter\@gobble\string#1@presq@nopostsq\endcsname[{##1}]{##2}}}
\expandafter\def\csname\expandafter\@gobble\string#1@nopresq@nopostsq\endcsname[##1]##2{#2}
\expandafter\def\csname\expandafter\@gobble\string#1@presq@nopostsq\endcsname[##1]##2{#3}
\expandafter\def\csname\expandafter\@gobble\string#1@nopresq@postsq\endcsname[##1]##2[##3]{#4}
\expandafter\def\csname\expandafter\@gobble\string#1@presq@postsq\endcsname[##1]##2[##3]{#5}
}

\newcommand{\NewcommandTwoArgsLastOpt}[3]
{
\DeclareRobustCommand#1{\csname\expandafter\@gobble\string#1@def\endcsname}
\expandafter\def\csname\expandafter\@gobble\string#1@def\endcsname##1{\@ifnextchar[%
{\csname\expandafter\@gobble\string#1@postsq\endcsname{##1}}%
{\csname\expandafter\@gobble\string#1@nopostsq\endcsname{##1}}}
\expandafter\def\csname\expandafter\@gobble\string#1@nopostsq\endcsname##1{#2}
\expandafter\def\csname\expandafter\@gobble\string#1@postsq\endcsname##1[##2]{#3}
}

\newcommand{\NewcommandFourArgsTwoOpt}[5]
{
\DeclareRobustCommand#1{\@ifnextchar[%
{\csname\expandafter\@gobble\string#1@presq\endcsname}%
{\csname\expandafter\@gobble\string#1@nopresq\endcsname}}
\expandafter\def\csname\expandafter\@gobble\string#1@nopresq\endcsname##1##2{\@ifnextchar[%
{\csname\expandafter\@gobble\string#1@nopresq@postsq\endcsname[]{##1}{##2}}%
{\csname\expandafter\@gobble\string#1@nopresq@nopostsq\endcsname[]{##1}{##2}}}
\expandafter\def\csname\expandafter\@gobble\string#1@presq\endcsname[##1]##2##3{\@ifnextchar[%
{\csname\expandafter\@gobble\string#1@presq@postsq\endcsname[{##1}]{##2}{##3}}%
{\csname\expandafter\@gobble\string#1@presq@nopostsq\endcsname[{##1}]{##2}{##3}}}
\expandafter\def\csname\expandafter\@gobble\string#1@nopresq@nopostsq\endcsname[##1]##2##3{#2}
\expandafter\def\csname\expandafter\@gobble\string#1@presq@nopostsq\endcsname[##1]##2##3{#3}
\expandafter\def\csname\expandafter\@gobble\string#1@nopresq@postsq\endcsname[##1]##2##3[##4]{#4}
\expandafter\def\csname\expandafter\@gobble\string#1@presq@postsq\endcsname[##1]##2##3[##4]{#5}
}

\newcommand{\NewcommandFourArgsTwoOptDefa}[4]
{
\expandafter\newcommand\csname\expandafter\@gobble\string#1@full\endcsname[4]{#4}
\NewcommandFourArgsTwoOpt{#1}
{\csname\expandafter\@gobble\string#1@full\endcsname{#2}{##2}{##3}{#3}}
{\csname\expandafter\@gobble\string#1@full\endcsname{##1}{##2}{##3}{#3}}
{\csname\expandafter\@gobble\string#1@full\endcsname{#2}{##2}{##3}{##4}}
{\csname\expandafter\@gobble\string#1@full\endcsname{##1}{##2}{##3}{##4}}
}

\NewcommandFourArgsTwoOpt{\ifndef}
{\@ifundefined{#2}{#3}{}}  %
{\ClassError{My Macros}{The first optional argument is not expected in "\ifndef"}{}}  %
{\@ifundefined{#2}{#3}{#4}}  %
{\ClassError{My Macros}{The first optional argument is not expected in "\ifndef"}{}}  %

\NewcommandThreeArgsTwoOpt{\myusepackage}
{\@ifpackageloaded{#2}{}{\usepackage{#2}}}
{\@ifpackageloaded{#2}{}{\usepackage[#1]{#2}}}
{\@ifpackageloaded{#2}{}{#3 \usepackage{#2}}}
{\@ifpackageloaded{#2}{}{#3 \usepackage[#1]{#2}}}

\makeatother

\myusepackage{amssymb}
\myusepackage{amsmath}  %

\myusepackage{amsthm}[ ]  %

\myusepackage{tempora}  %
\myusepackage[T1,OT2,T2A]{fontenc}

\usepackage[shorthands=off,greek,czech,ukrainian,russian,english]{babel}

\let\sq\relax  %

\DeclareTextSymbolDefault{\CYRYAT}{OT2}
\DeclareTextSymbolDefault{\cyryat}{OT2}
\DeclareTextSymbolDefault{\CYRFITA}{OT2}
\DeclareTextSymbolDefault{\cyrfita}{OT2}
\DeclareTextSymbolDefault{\CYRIZH}{OT2}
\DeclareTextSymbolDefault{\cyrizh}{OT2}

\myusepackage[utf8]{inputenc}
\let\f\relax

\myusepackage{soulutf8}      %
\setul{1.25pt}{.16pt}  %

\myusepackage{latexsym}

\myusepackage{amsfonts}

\myusepackage{units}    %
\newcommand{\dr}{\nicefrac}

\myusepackage[nolabel]{fnbreak}  %

\myusepackage{psfrag}   %

\myusepackage[hyphens,spaces,obeyspaces]{url}   %

\myusepackage{hyperref}  %
\setpdflinkmargin{0.75pt}  %

\myusepackage[usenames,dvipsnames]{xcolor}
\hypersetup{
colorlinks,
linkcolor={red!55!black},
citecolor={blue!55!black},
urlcolor={blue!35!black}
}

\myusepackage[linewidth=1pt]{mdframed}

\myusepackage{hyphenat}  %

\myusepackage{caption}   %

\myusepackage{sansmathfonts}  %

\myusepackage{microtype}   %

\myusepackage{marginnote}  %

\myusepackage{calc}  %

\myusepackage{mathrsfs}   %

\myusepackage{enumitem}  %

\myusepackage{graphicx}

\newcommand{\dgCapDefinition}{Definition}
\newcommand{\dgCapDefinitions}{Definitions}
\newcommand{\dgCapPostulate}{Postulate}
\newcommand{\dgCapPostulates}{Postulates}
\newcommand{\dgCapExample}{Example}

\newcommand{\dgCapFact}{Fact}
\newcommand{\dgCapFacts}{Facts}
\newcommand{\dgCapQuestion}{Question}
\newcommand{\dgCapQuestions}{Questions}
\newcommand{\dgCapLemma}{Lemma}
\newcommand{\dgCapLemmas}{Lemmas}
\newcommand{\dgCapNotation}{Notation}
\newcommand{\dgCapCorollary}{Corollary}
\newcommand{\dgCapCorollaries}{Corollaries}
\newcommand{\dgCapProposition}{Proposition}
\newcommand{\dgCapPropositions}{Propositions}
\newcommand{\dgCapClaim}{Claim}
\newcommand{\dgCapClaims}{Claims}
\newcommand{\dgCapTheorem}{Theorem}
\newcommand{\dgCapTheorems}{Theorems}
\newcommand{\dgCapProblem}{Problem}
\newcommand{\dgCapProblems}{Problems}
\newcommand{\dgCapRemark}{Remark}
\newcommand{\dgCapRemarks}{Remarks}
\newcommand{\dgCapConjecture}{Conjecture}
\newcommand{\dgCapConjectures}{Conjectures}
\newcommand{\dgCapResult}{Result}

\newcommand{\dgCapPart}{Part}
\newcommand{\dgCapParts}{Parts}
\newcommand{\dgCapChapter}{Chapter}
\newcommand{\dgCapChapters}{Chapters}
\newcommand{\dgCapSection}{Section}
\newcommand{\dgCapSections}{Sections}
\newcommand{\dgCapSubsection}{Subsection}
\newcommand{\dgCapSubsections}{Subsections}
\newcommand{\dgCapFigure}{Figure}
\newcommand{\dgCapFigures}{Figures}

\newcommand{\dgProofOf}{\proofname\ of}

\newcommand{\dgDefinition}{Definition}
\newcommand{\dgDefinitions}{Definitions}
\newcommand{\dgPostulate}{Postulate}
\newcommand{\dgPostulates}{Postulates}

\newcommand{\dgFact}{Fact}
\newcommand{\dgFacts}{Facts}
\newcommand{\dgQuestion}{Question}
\newcommand{\dgQuestions}{Questions}
\newcommand{\dgLemma}{Lemma}
\newcommand{\dgLemmas}{Lemmas}
\newcommand{\dgCorollary}{Corollary}
\newcommand{\dgCorollaries}{Corollaries}
\newcommand{\dgProposition}{Proposition}
\newcommand{\dgPropositions}{Propositions}
\newcommand{\dgClaim}{Claim}
\newcommand{\dgClaims}{Claims}
\newcommand{\dgTheorem}{Theorem}
\newcommand{\dgTheorems}{Theorems}
\newcommand{\dgProblem}{Problem}
\newcommand{\dgProblems}{Problems}
\newcommand{\dgRemark}{Remark}
\newcommand{\dgRemarks}{Remarks}
\newcommand{\dgConjecture}{Conjecture}
\newcommand{\dgConjectures}{Conjectures}

\newcommand{\dgPart}{Part}
\newcommand{\dgParts}{Parts}
\newcommand{\dgChapter}{Chapter}
\newcommand{\dgChapters}{Chapters}
\newcommand{\dgSection}{Section}
\newcommand{\dgSections}{Sections}
\newcommand{\dgSubsection}{Subsection}
\newcommand{\dgSubsections}{Subsections}
\newcommand{\dgFigure}{Figure}
\newcommand{\dgFigures}{Figures}

\newcommand{\dgFootnote}{Footnote}
\newcommand{\dgFootnotes}{Footnotes}

\ifndef{theorem}{}
\ifndef{theorem*}{\newtheorem*{theorem*}{\dgCapTheorem}}

\ifndef{lemma}{}
\ifndef{lemma*}{\newtheorem*{lemma*}{\dgCapLemma}}
\ifndef{corollary}{}
\ifndef{corollary*}{\newtheorem*{corollary*}{\dgCapCorollary}}
\ifndef{conjecture}{}
\ifndef{conjecture*}{\newtheorem*{conjecture*}{\dgCapConjecture}}
\ifndef{proposition}{}
\ifndef{proposition*}{\newtheorem*{proposition*}{\dgCapProposition}}
\ifndef{claim}{}
\ifndef{claim*}{\newtheorem*{claim*}{\dgCapClaim}}
\ifndef{result}{}
\ifndef{result*}{\newtheorem*{result*}{\dgCapResult}}
\ifndef{problem}{}
\ifndef{problem*}{\newtheorem*{problem*}{\dgCapProblem}}

\newtheoremstyle{mydefinition}  %
{\topsep}{\topsep}  %
{\slshape}  %
{}  %
{\bfseries}  %
{.}  %
{ }  %
{}  %

\newtheoremstyle{mynotation}  %
{\topsep}{\topsep}  %
{}  %
{}  %
{\bfseries\itshape}  %
{.}  %
{ }  %
{}  %

\newtheoremstyle{myremark}  %
{\topsep}{\topsep}  %
{\slshape}  %
{}  %
{\bfseries\itshape}  %
{.}  %
{ }  %
{\thmname{#1}\thmnumber{~#2}}  %

\newtheoremstyle{myexample}  %
{\topsep}{\topsep}  %
{\itshape}  %
{}  %
{\slshape}  %
{.}  %
{ }  %
{\underline{\thmname{#1}\thmnumber{~#2}}}  %

\newtheoremstyle{myclaims}  %
{\topsep}{\topsep}  %
{\slshape}  %
{}  %
{\bfseries\slshape}  %
{.}  %
{ }  %
{\thmname{#1}\thmnumber{~#2}\thmnote{\textnormal{~(#3)}}}  %

\ifndef{notation}{\theoremstyle{mynotation}}

{\theoremstyle{myremark}

\newtheorem*{myremark*}{\dgCapRemark}
}
{\theoremstyle{mydefinition}
\ifndef{postulate}{}
\ifndef{postulate*}{\newtheorem*{postulate*}{\dgCapPostulate}}
\ifndef{definition}{}
\ifndef{definition*}{\newtheorem*{definition*}{\dgCapDefinition}}
}
{\theoremstyle{myexample}
\ifndef{example}{}
\ifndef{example*}{\newtheorem*{example*}{\dgCapExample}}
}
{\theoremstyle{myclaims}

\newtheorem*{my_claim*}{\dgCapClaim}
\ifndef{fact}{}
\ifndef{fact*}{\newtheorem*{fact*}{\dgCapFact}}
\ifndef{question}{}
\ifndef{question*}{\newtheorem*{question*}{\dgCapQuestion}}
}

\newtheoremstyle{anystatementst}  %
{\topsep}{\topsep}  %
{\itshape}  %
{}  %
{\bfseries}  %
{.}  %
{ }  %
{#3}  %

{\theoremstyle{anystatementst} }

\newcommand{\newident}[3][\MyUniPat]{\ifthenelse{\equal{\MyUniPat}{#1}}
{
\newcommand{#2}[1][]{\ensuremath{\mathit{#3##1}}}
}
{\ifthenelse{\equal{}{#1}}
{
\newcommand{#2}[1][]{\ensuremath{\mathit{#3}}}
}
{
\DeclareRobustCommand{#2}[1][\MyUniPat]{\ifthenelse{\equal{\MyUniPat}{##1}}%
{\ensuremath{\mathit{#1}}}%
{\ensuremath{\mathit{#3}}}}
}
}
}

\newcommand{\newidenT}[3][\MyUniPat]{\ifthenelse{\equal{\MyUniPat}{#1}}
{
\DeclareRobustCommand{#2}[1][\MyUniPat]{\ifthenelse{\equal{\MyUniPat}{##1}}%
{\il{#3}}%
{\ensuremath{\mathit{#3##1}}}}
}
{
\DeclareRobustCommand{#2}[1][\MyUniPat]{\ifthenelse{\equal{\MyUniPat}{##1}}%
{\il{#1}}%
{\ensuremath{\mathit{#3}}}}
}
}

\newcommand{\newmat}[3][\MyUniPat]{\ifthenelse{\equal{\MyUniPat}{#1}}%
{\newcommand{#2}[1][]{#3##1}}%
{\newcommand{#2}[1][]{#3}}%
}

\newcommand{\providemat}[3][\MyUniPat]{\ifthenelse{\equal{\MyUniPat}{#1}}
{\providecommand{#2}[1][]{#3##1}}
{\providecommand{#2}[1][]{#3}}  %
}

\newcommand{\newmatop}[3][\MyUniPat]{\ifthenelse{\equal{\MyUniPat}{#1}}
{
\newcommand{#2}{\operatorname{#3}}
}
{
\DeclareRobustCommand{#2}[1][\MyUniPat]{\ifthenelse{\equal{\MyUniPat}{##1}}%
{\operatorname{#1}}%
{\operatorname{#3}}}
}
}

\newcommand{\newmatoparg}[3][\MyUniPat]{\ifthenelse{\equal{\MyUniPat}{#1}}
{
\DeclareRobustCommand{#2}[1]{\ifthenelse{\equal{}{##1}}{\operatorname{#3}}{\operatorname{#3}\lf(##1\rt)}}
}
{
\DeclareRobustCommand{#2}[2][\MyUniPat]{\ifthenelse{\equal{\MyUniPat}{##1}}%
{\ifthenelse{\equal{}{##2}}{\operatorname{#1}}{\operatorname{#1}\lf(##2\rt)}}%
{\ifthenelse{\equal{}{##2}}{\operatorname{#3}}{\operatorname{#3}\lf(##2\rt)}}}
}
}

\newcommand{\newOlike}[2]{
\DeclareRobustCommand{#1}[2][\MyUniPat]{\ifthenelse{\equal{\MyUniPat}{##1}}%
{\ensuremath{\mathit{#2}\lf(##2\rt)}}%
{#2(##2)}%
}}

\makeatletter
\newcommand{\MyMakeTheoMacros}[3]{
\expandafter\newcommand\csname\expandafter\@gobble\string#2NostarNoname@DGaux\endcsname[2][]
{\ifthenelse{\equal{}{##1}}%
{\begin{#1}~##2 \end{#1}}%
{\begin{#1}\label{##1}~##2\end{#1}}%
}
\expandafter\newcommand\csname\expandafter\@gobble\string#2StarNoname@DGaux\endcsname[1]
{\begin{#1*}~##1 \end{#1*}}
\newcommand#2{\expandafter\@ifstar%
\expandafter{\csname\expandafter\@gobble\string#2StarNoname@DGaux\endcsname}%
{\csname\expandafter\@gobble\string#2NostarNoname@DGaux\endcsname}%
}

\expandafter\newcommand\csname\expandafter\@gobble\string#2NostarName@DGaux\endcsname[3][]
{\ifthenelse{\equal{}{##1}}%
{\begin{#1}[\e{##2}]~##3 \end{#1}}%
{\begin{#1}[\e{##2}]\label{##1}~##3\end{#1}}%
}
\expandafter\newcommand\csname\expandafter\@gobble\string#2StarName@DGaux\endcsname[2]
{\begin{#1*}[\e{##1}]~##2 \end{#1*}}
\newcommand#3{\expandafter\@ifstar%
\expandafter{\csname\expandafter\@gobble\string#2StarName@DGaux\endcsname}
{\csname\expandafter\@gobble\string#2NostarName@DGaux\endcsname}%
}
}

\newtheorem*{rep@theorem}{\rep@title}
\newcommand{\newreptheorem}[2]{%
\newenvironment{rep#1}[1]{%
\def\rep@title{#2 \ref{##1}}%
\begin{rep@theorem}}%
{\end{rep@theorem}}}
\makeatother

\newcommand{\MyMakeDupTheoMacros}[7]{
\MyMakeTheoMacros{#1}{#2}{#3}
\newreptheorem{#1}{#6}
\newcommand{#4}[3]{
\newcommand{##2}{##3}
\begin{#1}\label{##1}~##2\end{#1}}
\newcommand{#5}[4]{
\newcommand{##2}{##4}
\begin{#1}{\e{##3}}\label{##1}~##2\end{#1}}
\newcommand{#7}[2]{\begin{rep#1}{##1}~##2 \end{rep#1}}
}

\newcommand{\MyMakeRefMacros}[3]{\newcommand{#1}{\dgref{#2}{#3}}}

\MyMakeTheoMacros{fact}{\fct}{\nfct}

\MyMakeRefMacros{\fctref}{\dgFact~}{\dgFacts~}
\MyMakeRefMacros{\Fctref}{\dgCapFact~}{\dgCapFacts~}

\MyMakeTheoMacros{question}{\quest}{\nquest}

\MyMakeRefMacros{\questref}{\dgQuestion~}{\dgQuestions~}
\MyMakeRefMacros{\Questref}{\dgCapQuestion~}{\dgCapQuestions~}

\MyMakeTheoMacros{notation}{\nota}{\nnota}

\MyMakeDupTheoMacros{lemma}
{\lem}{\nlem}{\lemdup}{\nlemdup}{\dgLemma}{\lemrep}

\MyMakeRefMacros{\lemref}{\dgLemma~}{\dgLemmas~}
\MyMakeRefMacros{\Lemref}{\dgCapLemma~}{\dgCapLemmas~}

\MyMakeDupTheoMacros{corollary}
{\crl}{\ncrl}{\crldup}{\ncrldup}{\dgCorollary}{\crlrep}

\MyMakeRefMacros{\crlref}{\dgCorollary~}{\dgCorollaries~}
\MyMakeRefMacros{\Crlref}{\dgCapCorollary~}{\dgCapCorollaries~}

\MyMakeTheoMacros{proposition}{\prp}{\nprp}

\newtheorem*{prp*}{\e{\dgCapProposition}}

\MyMakeRefMacros{\prpref}{\dgProposition~}{\dgPropositions~}
\MyMakeRefMacros{\Prpref}{\dgCapProposition~}{\dgCapPropositions~}

\MyMakeDupTheoMacros{my_claim}
{\clm}{\nclm}{\clmdup}{\nclmdup}{\dgClaim}{\clmrep}

\MyMakeRefMacros{\clmref}{\dgClaim~}{\dgClaims~}
\MyMakeRefMacros{\Clmref}{\dgCapClaim~}{\dgCapClaims~}

\MyMakeDupTheoMacros{theorem}
{\theo}{\ntheo}{\theodup}{\ntheodup}{\dgTheorem}{\theorep}

\MyMakeRefMacros{\theoref}{\dgTheorem~}{\dgTheorems~}
\MyMakeRefMacros{\Theoref}{\dgCapTheorem~}{\dgCapTheorems~}

\MyMakeTheoMacros{postulate}{\postu}{\npostu}

\MyMakeRefMacros{\posturef}{\dgPostulate~}{\dgPostulates~}
\MyMakeRefMacros{\Posturef}{\dgCapPostulate~}{\dgCapPostulates~}

\MyMakeTheoMacros{definition}{\defi}{\ndefi}

\MyMakeRefMacros{\defiref}{\dgDefinition~}{\dgDefinitions~}
\MyMakeRefMacros{\Defiref}{\dgCapDefinition~}{\dgCapDefinitions~}

\MyMakeTheoMacros{problem}{\prob}{\nprob}

\MyMakeRefMacros{\probref}{\dgProblem~}{\dgProblems~}
\MyMakeRefMacros{\Probref}{\dgCapProblem~}{\dgCapProblems~}

\MyMakeTheoMacros{myremark}{\rem}{\nrem}

\MyMakeRefMacros{\remref}{\dgRemark~}{\dgRemarks~}
\MyMakeRefMacros{\Remref}{\dgCapRemark~}{\dgCapRemarks~}

\MyMakeTheoMacros{conjecture}{\conj}{\nconj}

\MyMakeRefMacros{\conjref}{\dgConjecture~}{\dgConjectures~}
\MyMakeRefMacros{\Conjref}{\dgCapConjecture~}{\dgCapConjectures~}

\renewcommand{\qedsymbol}{$\blacksquare$}

\newcommand{\prfstart}[1][]{\ifthenelse{\equal{}{#1}}%
{\begin{proof}\renewcommand{\qedsymbol}{$\blacksquare$}}%
{\begin{proof}[\dgProofOf\ #1]%
\renewcommand{\qedsymbol}{$\blacksquare_{\mbox{\it{\scriptsize{#1}}}}$}}%
}
\newcommand{\prfend}[1][*]{%
\ifthenelse{\equal{}{#1}}{\renewcommand{\qedsymbol}{$\blacksquare$}}{}%
\ifthenelse{\equal{*}{#1}}{}%
{\renewcommand{\qedsymbol}{$\blacksquare_{\mbox{\it{\scriptsize{#1}}}}$}}%
\end{proof}\renewcommand{\qedsymbol}{$\blacksquare$}%
}

\newcommand{\NewSectLikeDG}[3]{
\NewcommandThreeArgsTwoOpt{#1}
{\ifthenelse{\equal{*}{##2}}{#2*}{#2{##2}}}
{#2{##2\label{##1}}}
{#2[##3]{##2#3{##3}}#3{##3}}
{#2[##3]{##2#3{##3}\label{##1}}#3{##3}}
}

\NewSectLikeDG{\sect}{\section}{\sectionmark}

\NewSectLikeDG{\ssect}{\subsection}{\subsectionmark}

\NewSectLikeDG{\sssect}{\subsubsection}{\subsubsectionmark}

\makeatletter
\@addtoreset{chapter}{part}
\makeatother

\newcommand*\parttitle{}
\let\origpart\part
\renewcommand*{\part}[2][]{%
\ifx\\#1\\
\origpart{#2}%
\renewcommand*\parttitle{#2}%
\else
\origpart[#1]{#2}%
\renewcommand*\parttitle{#1}%
\fi
}

\ifndef{chapter}
{\NewSectLikeDG{\chap}{\part}{\DoNothing}}  %
[\NewSectLikeDG{\chap}{\chapter}{\chaptermark}]

\NewSectLikeDG{\prt}{\part}{\DoNothing}  %

\newcommand{\para}[2][]{\ifthenelse{\equal{}{#1}}
{\paragraph{#2}}
{\paragraph{#2\label{#1}}}}

\NewcommandTwoArgsLastOpt{\parass}
{\para{#1}\toc{subsection}{#1}}
{\para{#1#2}\toc{subsection}{#1}}

\makeatletter

\def\@seccntformat#1{\@ifundefined{#1@dgformat}%
{\csname the#1\endcsname\quad}%
{\csname #1@dgformat\endcsname.~}}%

\renewcommand\theparagraph{\arabic{paragraph}}
\newcommand{\paragraph@dgformat}{\S\theparagraph}

\renewcommand\thesubparagraph{\arabic{paragraph}.\arabic{subparagraph}}
\newcommand{\subparagraph@dgformat}{\S\thesubparagraph}

\MyMakeRefMacros{\paref}{\S}{\S\S~}
\makeatother

\MyMakeRefMacros{\prtref}{\dgPart~}{\dgParts~}
\MyMakeRefMacros{\Prtref}{\dgCapPart~}{\dgCapParts~}

\MyMakeRefMacros{\chref}{\dgChapter~}{\dgChapters~}
\MyMakeRefMacros{\Chref}{\dgCapChapter~}{\dgCapChapters~}

\MyMakeRefMacros{\sref}{\dgSection~}{\dgSections~}
\MyMakeRefMacros{\Sref}{\dgCapSection~}{\dgCapSections~}

\MyMakeRefMacros{\ssref}{\dgSubsection~}{\dgSubsections~}
\MyMakeRefMacros{\Ssref}{\dgCapSubsection~}{\dgCapSubsections~}

\MyMakeRefMacros{\sssref}{\dgSubsection~}{\dgSubsections~}
\MyMakeRefMacros{\Sssref}{\dgCapSubsection~}{\dgCapSubsections~}

\newcommand{\toc}[3][]{%
\ifthenelse{\equal{chapter}{#2}}{\newpage}{}%
\ifthenelse{\equal{part}{#2}}{\newpage}{}%
\ifthenelse{\equal{}{#1}}%
{\phantomsection\addcontentsline{toc}{#2}{#3}}%
{\phantomsection\refstepcounter{#2}\label{#1}\addcontentsline{toc}{#2}{#3}}%
}

\newcommand{\toct}[1]{\toc{section}{#1}}

\definecolor{DarkRed}{rgb}{0.65,0.05,0.05}

\definecolor{LightRed}{rgb}{0.855,0.16,0.16}

\definecolor{DarkGreen}{rgb}{0, 0.52, 0.05}

\definecolor{LightGreen}{rgb}{0.08,0.855,0.16}

\definecolor{DarkBlue}{rgb}{0.05, 0, 0.55}

\definecolor{LightBlue}{rgb}{0.122,0.016,0.855}

\MyMakeRefMacros{\figref}{\dgFigure~}{\dgFigures~}
\MyMakeRefMacros{\Figref}{\dgCapFigure~}{\dgCapFigures~}

\newcommand{\IfMathMode}[2]{\ifmmode{#1}\else{#2}\fi}

\newcommand{\fbr}[1]{
\delimiterfactor=1001 %
\IfMathMode%
{#1}{$#1$}\delimiterfactor=901%
}

\newcommand{\fnbr}[1]{\mbox{\fbr{#1}}}  %

\newcommand{\f}[2][*]{\ifthenelse{\equal{}{#1}}{\fbr{#2}}{\fnbr{#2}}}

\newcommand{\mal}[2][]{\MyChangeMathMargins
\delimiterfactor=1001 %
\ifthenelse{\equal{}{#1}}%
{\begin{align*} #2 \end{align*}}%
{\ifthenelse{\equal{P}{#1}}%
{\allowdisplaybreaks\begin{align*} #2%
\end{align*}\interdisplaylinepenalty=10000}%
{\begin{align}\begin{split}\label{#1} #2 \end{split}\end{align}}%
}\delimiterfactor=901%
}

\newcommand{\m}{\mal}

\makeatletter

\DeclareRobustCommand\bref{\@ifnextchar[%
{\bref@presq}%
{\bref@nopresq}%
}

\def\bref@presq[#1]{\@ifnextchar[%
{(\ref{#1}), \bref@presq}
{(\ref{#1}) and~\bref@nopresq}
}

\def\bref@nopresq#1{(\ref{#1})}

\DeclareRobustCommand\dgref[2]{\@ifnextchar[%
{#2\dgref@presq}%
{#1\dgref@nopresq}%
}

\def\dgref@presq[#1]{\@ifnextchar[%
{\ref{#1}, \dgref@presq}
{\ref{#1} and~\dgref@nopresq}
}

\def\dgref@nopresq#1{\ref{#1}}

\newcommand\Cases[1][0pt]{%
\def\Case@skipam{#1}%
\left\{\!\!\!\begin{array}{ll}\Cases@continue%
}

\def\Cases@continue#1#2{\@ifnextchar\bgroup%
{#1 &\txt{#2}\\[\Case@skipam] \Cases@continue}%
{#1 &\txt{#2}\end{array}\Cases@end}%
}

\def\Cases@end{\@ifnextchar[%
{\Cases@end@postsq}%
{\right.}
}

\def\Cases@end@postsq[#1]{\ifthenelse{\equal{\}}{#1}}
{\!\!\right\}}%
{\right.}
}

\makeatother

\newcommand{\lf}{\mathopen{}\mathclose\bgroup\left}
\newcommand{\rt}{\aftergroup\egroup\right}

\providecommand{\middle}{\big}
\newcommand{\md}{\middle}

\newcommand{\ud}{\vphantom{|_1^1}}

\newcommand{\chs}{\genfrac(){0cm}{}}  %

\newmatop{\plog}{poly-log}

\newmatoparg{\supp}{supp}   %
\newmatoparg{\sgn}{sgn}     %
\newmatoparg{\diam}{diam}   %
\newmatoparg{\Span}{span}   %
\newmatoparg{\Dim}{dim}     %
\newmatoparg{\pow}{pow}

\newcommand{\NewHLikeDG}[2]{
\NewcommandThreeArgsTwoOpt{#1}
{\operatorname{\mathnormal{#2}}\lf(##2\rt)}
{\operatorname{\mathnormal{#2_{##1}}}\lf(##2\rt)}
{\operatorname{\mathnormal{#2}}\lf(##2\md|{\ud}##3\rt)}
{\operatorname{\mathnormal{#2_{##1}}}\lf(##2\md|{\ud}##3\rt)}
}

\NewHLikeDG{\h}{H}

\NewHLikeDG{\hm}{H_{min}}

\newcommand{\NewILikeDG}[2]{
\NewcommandFourArgsTwoOpt{#1}
{\mathop{\pmb{#2}}\lf[##2:{\ud}##3\rt]}
{\mathop{\pmb{#2}\?\!_{##1}}\lf[##2:{\ud}##3\rt]}
{\mathop{\pmb{#2}}\lf[##2:##3\md|{\ud}##4\rt]}
{\mathop{\pmb{#2}\?\!_{##1}}\lf[##2:##3\md|{\ud}##4\rt]}
}

\NewILikeDG{\I}{I}

\newcommand{\NewELikeDG}[2]{
\NewcommandThreeArgsTwoOpt{#1}
{#2\lf[##2\rt]}
{#2_{##1}\lf[##2\rt]}
{#2\lf[##2\md|{\ud}##3\rt]}
{#2_{##1}\lf[##2\md|{\ud}##3\rt]}
}

\NewELikeDG{\PR}{\mathop{\pmb{Pr}}}

\NewELikeDG{\E}{\mathop{\pmb{E}}}

\NewELikeDG{\Del}{\mathop{\pmb{\Delta}}}

\NewELikeDG{\Var}{\mathop{\pmb{Var}}}

\NewcommandTwoArgsLastOpt{\GF}
{{\mathcal{G\!F}_{#1}}}
{{\mathcal{G\!F}_{#1}^{#2}}}

\providemat{\NN}{\mathbb{N}}
\providemat{\RR}{\mathbb{R}}

\newcommand{\wtl}{\widetilde}

\newcommand{\pss}[1][]{\ifthenelse{\equal{}{#1}}%
{\txt{'s}}%
{\f{#1\txt{'s}}}}

\newcommand{\pl}[1][]{\ifthenelse{\equal{}{#1}}%
{\mskip-6mu\stackrel{\text-}{}\mskip-4mu\txt{s}}%
{\f{#1\mskip-6mu\stackrel{\text-}{}\mskip-4mu\txt{s}}}}

\newcommand{\ord}[1][]{\ifthenelse{\equal{}{#1}}%
{\txt{'th}}%
{\ifthenelse{\equal{1}{#1}}{$1\txt{'st}$}{\ifthenelse{\equal{2}{#1}}{$2\txt{'nd}$}{\ifthenelse{\equal{3}{#1}}{$3\txt{'rd}$}{\f{#1\txt{'th}}}}}}}

\newcommand{\fr}[3][*]{%
\ifthenelse{\equal{*}{#1}}%
{\frac{#2}{#3}}{}%
\ifthenelse{\equal{}{#1}}%
{\dr{#2}{#3}}{}%
\ifthenelse{\equal{/}{#1}}%
{\lf.#2\md/#3\rt.}{}%
\ifthenelse{\equal{p_}{#1}}%
{\lf.\lf(#2\rt)\md/#3\rt.}{}%
\ifthenelse{\equal{_p}{#1}}%
{\lf.#2\md/\lf(#3\rt)\rt.}{}%
\ifthenelse{\equal{pp}{#1}}%
{\lf.\lf(#2\rt)\md/\lf(#3\rt)\rt.}{}%
}

\newcommand{\sq}{\sqrt}

\NewcommandThreeArgsTwoOpt{\set}
{\lf\{#2\rt\}}
{}
{\lf\{#2\md|\ud#3\rt\}}
{\lf\{#2\ud#1\ud#3\rt\}}

\newcommand{\Log}[2][]{\ifthenelse{\equal{}{#1}}%
{\log\lf(#2\rt)}%
{\log_{#1}\lf(#2\rt)}%
}

\newcommand{\NewMinLikeDG}[2]{
\NewcommandThreeArgsTwoOpt{#1}
{#2\lf\{##2\rt\}}
{#2_{##1}\lf\{##2\rt\}}
{#2\lf\{##2\md|\ud##3\rt\}}
{#2_{##1}\lf\{##2\md|\ud##3\rt\}}
}

\NewMinLikeDG{\Min}{\min}

\NewMinLikeDG{\Max}{\max}

\newmatop{\argmin}{argmin}
\NewMinLikeDG{\Argmin}{\argmin}

\newmatop{\argmax}{argmax}
\NewMinLikeDG{\Argmax}{\argmax}

\NewMinLikeDG{\Sup}{\sup}

\NewMinLikeDG{\Inf}{\inf}

\newOlike{\asO}{O}
\newOlike{\astO}{\wtl O}
\newOlike{\aso}{o}
\newOlike{\asOm}{\Omega}
\newOlike{\astOm}{\wtl\Omega}
\newOlike{\asom}{\omega}
\newOlike{\asT}{\Theta}
\newOlike{\astT}{\wtl\Theta}

\makeatletter

\DeclareRobustCommand\bk{\@ifnextchar[%
{\lf\langle \bk@bra}%
{\@ifnextchar<%
{\lf.\bk@op}
{\lf| \bk@ket}%
}%
}

\def\bk@bra[#1]{#1 \@ifnextchar[%
{\md|\hspace{-1.5pt}\md\langle \bk@bra}%
{\@ifnextchar\bgroup%
{\md| \bk@ket}%
{\@ifnextchar<%
{\md| \bk@op}%
{\rt|}%
}%
}%
}

\def\bk@ket#1{#1 \@ifnextchar[%
{\md\rangle\hspace{-1.5pt}\md\langle \bk@bra}%
{\@ifnextchar\bgroup%
{\md\rangle\hspace{-1.5pt}\md| \bk@ket}%
{\@ifnextchar<%
{\md\rangle \bk@op}%
{\rt\rangle}%
}%
}%
}

\def\bk@op<#1>{#1 \@ifnextchar[%
{\md\langle \bk@bra}%
{\@ifnextchar\bgroup%
{\md| \bk@ket}%
{\@ifnextchar<%
{\tm \bk@op}%
{\rt.}%
}%
}%
}

\makeatother

\NewcommandTwoArgsLastOpt{\sz}
{\lf|#1\rt|}
{\lf|#1\rt|_{#2}}

\NewcommandTwoArgsLastOpt{\norm}
{\lf\|#1\rt\|}%
{\lf\|#1\rt\|_{#2}}

\newcommand{\txt}[1]{\textrm{#1}}  %

\newcommand{\Cl}{\mathcal}   %

\newcommand{\Cel}{\mathscr}  %

\DeclareMathAlphabet{\mathbfcal}{OMS}{cmsy}{b}{n}

\DeclareMathAlphabet{\mathlowcal}{OT1}{pzc}{m}{it}

\newcommand{\XY}{{X\!\fromto\!Y}}   %
\newident{\Pxy}{P^\XY}
\newident{\UPxy}{UP^\XY}
\newident{\coNPxy}{coNP^\XY}
\newident{\NPxy}{NP^\XY}

\newcommand{\nin}{\not\in}  %

\newcommand{\fromto}{\leftrightarrow}
\newcommand{\Then}{\Longrightarrow}

\newcommand{\dt}{\cdot}
\newcommand{\tm}{\cdot}
\newcommand{\deq}{\stackrel{\textrm{def}}{=}}
\newcommand{\smin}{\setminus}
\newcommand{\asgn}{\pmb\leftarrowtail}  %

\newcommand{\sbseq}{\subseteq}
\newcommand{\sbs}{\subset}

\newcommand{\es}{\emptyset}

\newcommand{\bigland}{\bigwedge}
\newcommand{\biglor}{\bigvee}

\NewcommandFourArgsTwoOptDefa{\overlay}
{0mu}{1}
{
\begingroup
\mathchoice{
\bgroup
\scalebox{#4}{\ooalign{$\displaystyle#2\mkern#1$\cr
\hidewidth{$\displaystyle\mkern#1#3$}\hidewidth}}
\egroup
}{
\bgroup
\scalebox{#4}{\ooalign{$\textstyle#2\mkern#1$\cr
\hidewidth{$\textstyle\mkern#1#3$}\hidewidth}}
\egroup
}{
\bgroup
\scalebox{#4}{\ooalign{$\scriptstyle#2\mkern#1$\cr
\hidewidth{$\scriptstyle\mkern#1#3$}\hidewidth}}
\egroup
}{
\bgroup
\scalebox{#4}{\ooalign{$\scriptscriptstyle#2\mkern#1$\cr
\hidewidth{$\scriptscriptstyle\mkern#1#3$}\hidewidth}}
\egroup
}
\endgroup
}

\newcommand{\unin}{\mathrel{\overlay[1.25mu]{\subset}{\sim}}}

\providecommand{\hrectangle}{{\overlay[6mu]{\sqsubset}{\sqsupset}[0.625]}}

\makeatletter
\def\mov@rlay#1#2{\leavevmode\vtop{%
\baselineskip\z@skip \lineskiplimit-\maxdimen
\ialign{\hfil$\m@th#1##$\hfil\cr#2\crcr}}}
\makeatother

\newcommand{\ds}[1][]
{\ifthenelse{\equal{}{#1}}{\allowbreak\dots}{#1\allowbreak\dots#1}}
\newmat{\dc}{\ds[,]}

\newmat{\OI}{\set{0,1}} 

\newmat{\TF}{\set{\top,\bot}}  %

\mathchardef\myhyphen="2D

\DeclareRobustCommand{\g}[1]{%
\ifthenelse{\equal{#1}-}{\myhyphen}{}%
\ifthenelse{\equal{#1}=}{\equiv}{}%
}

\makeatletter

\let\dgampersand\&

\DeclareRobustCommand\&{%
\new@ifnextchar[%
{\dgsep@reposit}%
{\dgampersand}%
}

\def\dgsep@reposit[#1]{\hspace{#1}&\hspace{-#1}}

\makeatother

\newcommand{\abstart}{\begin{abstract}}
\newcommand{\abend}{\end{abstract}}

\newenvironment{myepig}
{\par\addtolength{\leftskip}{28mm}\addtolength{\rightskip}{8mm}\noindent\ignorespaces}
{\par}

\newenvironment{myepigsgn}
{\par\addtolength{\leftskip}{84mm}\noindent\ignorespaces}
{\par}

\NewcommandThreeArgsTwoOpt{\epig}
{{\small \begin{myepig} \textit{#2} \end{myepig} \addvspace{\baselineskip}}}
{\ClassError{My Macros}{The first optional argument is not expected in "\epig"}{}}
{{\small \begin{myepig} \textit{#2} \end{myepig} \Nopagebreak%
\begin{myepigsgn} -- #3 \end{myepigsgn}\addvspace{\baselineskip}}}
{\ClassError{My Macros}{The first optional argument is not expected in "\epig"}{}}

\newcommand{\itstart}[1][\MyUniPat]{%
\ifthenelse{\equal{\MyUniPat}{#1}}%
{\begin{itemize}[noitemsep,topsep=3pt]}%
{\ifthenelse{\equal{*}{#1}}%
{\begin{itemize}[noitemsep,topsep=3pt,leftmargin=*]}%
{\begin{itemize}[#1]}%
}%
}
\newcommand{\enstart}[1][\MyUniPat]{%
\ifthenelse{\equal{\MyUniPat}{#1}}%
{\begin{enumerate}[noitemsep,topsep=3pt]}%
{\ifthenelse{\equal{*}{#1}}%
{\begin{enumerate}[noitemsep,topsep=3pt,leftmargin=*]}%
{\begin{enumerate}[#1]}%
}%
}

\newcommand{\itend}{\end{itemize}}
\newcommand{\enend}{\end{enumerate}}

\newcommand{\DoNothing}[1]{}  %

\makeatletter

\protected \def \dg{\@ifstar\dg@st\dg@nost} %

\protected \def \dg@nost #1{%
\textcolor{Red}
{
{\normalmarginpar\marginnote{\bl{DG's note}}}
{\reversemarginpar\marginnote{\bl{DG's note}}\\}
\IfMathMode{
~~~\txt{#1}~
}{
~\\~~~#1~\\
{\normalmarginpar\marginnote{\bl{\ul{------}}}}
{\reversemarginpar\marginnote{\bl{\ul{------}}}\\}
}
}

}

\protected \def \dg@st #1{%
~\\\textcolor{Red}{%
$\blacktriangleleft\blacktriangleright$\\\\
#1\\\\
$\blacktriangleleft\blacktriangleright$}%
}

\makeatother

\newcommand{\fn}[2][]{%
\IfMathMode{}{}%
\ifthenelse{\equal{}{#1}}%
{\footnote{
\ignorespaces #2\unskip}}%
{\footnote{\label{#1}
\ignorespaces #2}}%
}

\newcommand{\fnm}{\footnotemark}
\newcommand{\fnt}[2][]{\ifthenelse{\equal{}{#1}}%
{\footnotetext{
\ignorespaces #2}}%
{\footnotetext{\label{#1}
\ignorespaces #2}}%
}

\MyMakeRefMacros{\fnref}{\dgFootnote~}{\dgFootnotes~}

\makeatletter

\newcommand{\Nopagebreak}{\par\nobreak\@afterheading} 

\makeatother

\DeclareTextFontCommand{\bemph}{\bfseries}
\DeclareTextFontCommand{\ibemph}{\bfseries\em}

\newcommand{\e}{\emph}

\newcommand{\ebi}{\ibemph}

\newcommand{\bl}[1]{{\bf #1}} %

\newcommand{\il}[1]{{\it #1}} %

\def\?{\mskip 1.5mu} %

\newcommand{\MyChangeMathMargins}{%
\setlength{\abovedisplayskip}{\abovedisplayshortskip + 4pt}%
\setlength{\belowdisplayskip}{\abovedisplayshortskip + 5pt}%
}

\newident{\FPxy}{FewP^\XY}
\newident{\FPtxy}{FewP_t^\XY}
\newident{\PNPrxy}{PNP_\hrectangle^\XY}
\newident{\PNPxy}{PNP^\XY}

\newident{\XiYtZ}{[(X \fromto Y) \to Z]}
\newident{\XYtZ}{[(X,Y) \to Z]}
\newident{\XZiYZ}{[(X,Z) \fromto (Y,Z)]}
\newcommand{\XYZcap}{$\XZiYZ \cap \XYtZ$}

\title{Patterned non-determinism in communication complexity\fn
{This is a preliminary version...}}

\newcommand{\instDG}{Institute of Mathematics of the Czech Academy of Sciences, \v Zitna 25, Praha 1, Czech Republic.}

\newcommand{\DmytroG}{In 2022 the author has changed the English spelling of his first name from the previous russian-odoured form \e{``Dmitry''} to the Ukrainian \e{``Dmytro''}.}

\newcommand{\thanksDG}{Partially funded by the grant 19-27871X of GA \v CR and by RVO:\ 67985840.
Part of this work was done while visiting the Centre for Quantum Technologies at the National University of Singapore.}

\author{Dmytro Gavinsky\thanks{\DmytroG} \thanks{\instDG\newline\thanksDG}
}

\begin{document}

\maketitle

\thispagestyle{empty}

\abstart

We define and study the model of \e{patterned non-determinism} in bipartite communication complexity, denoted by \PNPxy.
It generalises the known models \UPxy\ and \FPxy\ through relaxing the constraints on the witnessing structure of the underlying \NPxy-protocol.

It is shown that for the case of total functions \PNPxy\ equals \Pxy\ (similarly to \UPxy\ and \FPxy).
Moreover, the corresponding \e{exhaustive witness-searching} problem -- determining the \e{full set} of witnesses that lead to the acceptance of a given input pair -- also has an efficient deterministic protocol.

Structurally, the possibility of efficient exhaustive \PNPxy-search summarises the above results and can be stated like this:\ if $f_1\dc f_m$ are bipartite total Boolean functions with efficient deterministic protocols, then for every input $(x,y)$ the set $\set{i}[f_i(x,y)=\top]$ can be found by a deterministic protocol of cost \e{poly-logarithmic} in $n$ and the total number of such sets for these \pss[f_i].

Finally, the possibility of efficient exhaustive \PNPxy-search is used to analyse certain \e{three-party communication} regime (under the ``number in hand'' input partition):
The corresponding three-party model is shown to be as strong qualitatively as the weakest among its two-party amplifications obtained by allowing free communication between a pair of players.

\abend

\sect[s_intro]{Introduction}

Let $f(x,y):\: \OI^n\times \OI^n\to\TF$ be a total bipartite communication problem with an efficient \NPxy-protocol $\Pi$, that is, the total number of bits sent by $\Pi(x,y)$ is in $\plog(n)$.
It was shown by Yannakakis~\cite{Y91_Ex_Co} that if for every $(x,y)\in f^{-1}(\top)$ there is exactly one \f\Pi-witness, then $f\in \Pxy$.
Later Karchmer, Newman, Saks and Wigderson~\cite{KNSW94_Non} strengthened the result by drawing the same conclusion from the weaker assumption that the number of \f\Pi-witnesses per input pair was at most $\plog(n)$.
The corresponding communication complexity classes -- that is, the families of functions for which there are efficient protocols -- are denoted by \UPxy\ and \FPxy\ and the above results can be stated as $\UPxy = \FPxy = \Pxy$.\fn
{
To denote communication complexity \e{classes}, as well as the corresponding \e{models}, we will usually add the superscript ``$\XY$'' to the common notation for the corresponding computational complexity class.
}

Consider the following generalisation.
Let $\Cl W$ be the family of all possible \pss[\Pi] witnesses and call $\gamma\sbseq \Cl W$ a \e{pattern} (for the protocol $\Pi$) if for some $(x,y)\in f^{-1}(\top)$ the set of witnesses that cause \pss[\Pi] acceptance of $(x,y)$ equals $\gamma$.
If the total number of \pss[\Pi] patterns is at most $2^{k(n)}$ for $k(n) \in \plog(n)$, then we say that $\Pi$ is an efficient protocol in the model of \e{patterned non-determinism}, and the Boolean function $f$ that $\Pi$ computes belongs to the corresponding communication complexity class \PNPxy\ (obviously, $\UPxy \sbseq \FPxy \sbseq \PNPxy$).
We will see that $\PNPxy = \Pxy$.

Next we consider the communication complexity of the \e{exhaustive witness-search problem} corresponding to \PNPxy-protocols (or simply \e{exhaustive \PNPxy-search}):\ by this we will mean determining the \e{full set} of witnesses that lead to the acceptance of a given input pair by the given \PNPxy-protocol (which is, in particular, an \NPxy-protocol).

On the one hand, we will see that from the equality of a certain subclass $\Cel C\sbseq \NPxy$ to \Pxy\ even the possibility of efficiently finding \e{any protocol-compatible witness} doesn't follow in general (leave alone determining the complete set of valid witnesses).\fn
{
Even though $\Cel C \sbseq \Pxy$ trivially implies the existence of \e{some} efficiently verifiable witness for every answer to a problem $f(x,y) \in \Cel C$, it can be the case that witnessing in accordance with the same \NPxy-protocol that establishes the membership of $f$ in $\Cel C$ is not feasible:
We will see that there are subsets $\Cl A, \Cl B \sbseq \OI^n$ such that the (total) instance of the \e{set intersection problem} defined over $\Cl A\times \Cl B$ belongs to $\NPxy \cap \coNPxy$ -- therefore to \Pxy~\cite{AUY83_On} -- while at the same time finding a presumably existing index $i\in [n]$ such that $x_i=y_i=1$ is not only infeasible deterministically, but also hard for randomised protocols over the uniformly-random input from $\set{(x,y)\in \Cl A\times \Cl B}[\exists\: i:\: x_i=y_i=1]$.
}
Nevertheless, efficient deterministic witness-searching for \PNPxy-protocols will be presented.
That is, for an efficient \NPxy-protocol $\Pi$ with at most $2^{\plog(n)}$ patterns there exists a \Pxy-protocol $\Pi_{\txt{search}}$ of cost at most $\plog(n)$ that finds -- for every input pair $(x,y)$ that $\Pi$ accepts -- \e{the exact set of witnesses} that lead to \pss[\Pi] acceptance of $(x,y)$.\fn
{
Note that the set can, in general, be large, containing as many elements as there are different witnesses in $\Pi$.
}

More formally, if $f_1(x,y)\dc f_m(x,y):\: \OI^n\times \OI^n\to\TF$ are such that every $f_i$ has an efficient deterministic protocol and the set $\set{\set{i}[f_i(x,y)=\top]}_{x,y}$ is of size at most quasi-polynomial in $n$, then the exhaustive-search function $F(x,y)\deq \set{i}[f_i(x,y)=\top]$ is in \Pxy\ (i.e., there is an efficient deterministic protocol).
Note that this statement generalises the equality $\PNPxy = \Pxy$.

Finally, let $g(x,y,z):\: \OI^n\times \OI^n\times\OI^n\to\OI$ be a tripartite total function.
Consider the following scenarios of deterministically computing $g$:\,\fn[fn_notation]
{
The intuition behind the notation used next is clear:\ ``\XYtZ'' means that a holder of $X$ and $Y$ sends a 1-way message to the holder of $Z$, ``\XZiYZ'' means that a holder of $X$ and $Z$ is allowed to interact in the 2-way regime with the holder of $Y$ and $Z$ and so on.
}
\itstart
\item Denote by \XiYtZ\ the regime where Alice receives $x$, Bob receives $y$, Charlie receives $z$, Alice and Bob interact in order to produce a message that is sent to Charlie, who must answer upon receiving it (alternatively, this setting can be viewed as having ``broadcasting'' interaction between Alice and Bob, that is, letting Charlie see its transcript).
\item Denote by \XYtZ\ the regime where Alice receives $(x,y)$, Charlie receives $z$, Alice sends a message to Charlie, who must answer upon receiving it.
\item Denote by \XZiYZ\ the regime where Alice receives $(x,z)$, Bob receives $(y,z)$, they interact until Bob produces the answer.
\itend
For brevity we will say that \e{$g(x,y,z)$ is efficiently computable in \XYZcap} if it has efficient protocols in both \XZiYZ\ and \XYtZ.

Note that computing a tripartite function in \XiYtZ\ is at least as hard as computing it in \XYZcap:
Assume the existence of an efficient three-player protocol $\Pi$ in the setting \XiYtZ, then the existence of an efficient protocol in \XYtZ\ follows by ``merging'' Alice and Bob in $\Pi$ (i.e., letting them communicate for free) and an efficient protocol in \XZiYZ\ can be obtained by ``merging'' Bob and Charlie (the resulting protocol only uses even more restricted setting that can be denoted by $[X \fromto (Y,Z)]$, where Alice doesn't receive $Z$ as part of her input).

We will see that every total function $g(x,y,z)$ that is efficiently computable in \XYZcap\ has an efficient protocol in \XiYtZ\ as well:\ this will follow from the possibility of efficient exhaustive \PNPxy-search.
In particular, the model of \e{deterministic interactive Alice and Bob with listening Charlie}, \XiYtZ, is as strong qualitatively as the weakest among its two-party amplifications obtained by allowing free communication between a pair of players.

\ssect*{Related work}

For both \e{non-deterministic} and \e{randomised} setting in the \e{unrestricted} interactive three-party case, Draisma, Kushilevitz and Weinreb~\cite{DKW11_Part} have demonstrated an \e{exponential gap} between the communication complexity of a tripartite total function and the largest of its three bipartite complexities in the amplified models resulting from allowing free communication between a pair of players.
The new three-party result (\sref{s_three}) can be viewed as complementary to~\cite{DKW11_Part}:\ it shows that the gap is \e{at most polynomial} in the case of \e{deterministic} interactive Alice and Bob with \e{listening Charlie}.
The case of three \e{deterministic} players with \e{unrestricted} interaction remains open.

\sect[s_defi]{Preliminaries and definitions}

We will write $[n]$ to denote the set $\set{1\dc n}\sbs\NN$.
Let $(a,\, b)$, $[a,\, b]$, $[a,\, b)$ and $(a,\, b]$ denote the corresponding open, closed and half-open intervals in $\RR$.
For a finite $S\sbs\NN$ we will write $S(i)$ to address the \ord[i] element of $S$ in natural ordering.
For any set $S$ we will denote by $\pow S$ the family of its subsets and by $\chs St$ the family of size-$t$ subsets.
We will write $x\unin S$ to say that $x$ is a uniformly random element of $S$. 
Towards readability, we will allow both $\set{\dt}[\dt]$ and $\set[:]{\dt}[\dt]$ to denote sets with conditions (preferring the former).

For $x\in\OI^n$ and $i\in{[n]}$, we will write $x_i$ or $x(i)$ to address the \ord[i] bit of $x$ (preferring ``$x_i$'' unless it may cause ambiguity).
Let $\sz x$ denote the Hamming weight of $x$.
At times we will implicitly assume (without causing ambiguity) the trivial isomorphism between the \f n-bit strings and the subsets of $[n]$: in particular, the notation $\chs{[n]}k$ will stand for $\set{x\in\OI^n}[\sz{x}=k]$, and $x\cap y$ will address the set $\set{i\in[n]}[x_i=y_i=1]$.

Let $\bot$ and $\top$ denote, respectively, the false and the true values:\ sometimes we will use the Boolean domain $\set{\bot,\top}$ (instead of $\OI$) to emphasise the intuitive asymmetry between the two values (say, when the non-deterministic computation is distinguished from the co-non-deterministic one, or if there is a ``clear logical flavour'' inherent to the values).

By default the logarithms are base-$2$.

We will use $\asgn$ to denote the assignment operation (e.g., in algorithms).

\ssect[ss_CommC]{Communication complexity}

The study of communication complexity was initiated by Abelson~\cite{A78_Lo} in the regime of real-valued messages and adapted by Yao~\cite{Y79_So} to the discrete regime that we are interested in.
We refer the reader to~\cite{KN97_Comm} for a classical background on communication complexity in general, to~\cite{GPW18_The} for a great survey of the more recent structural developments and to~\cite{DKW11_Part} for some insight into the multi-party communication complexity setting.

Unless stated otherwise, the communication problems considered in this work are \e{total functions} (the only exception will be \e{witness-search problems}).

We will add the superscript ``$\XY$'' to the common notation for a computational complexity class to denote the corresponding communication complexity class (e.g., \Pxy\ or \NPxy).
The resulting symbol will be used in three ways:\ to address the class itself; to address the corresponding communication model; to denote the complexity of a communication problem in that model (e.g., $\Pxy(f)$ is the deterministic communication complexity of $f$).

As the standard models \Pxy\ and \NPxy\ are of core importance for this work, their definitions for the case of total functions are given next for the reader's convenience.

\ndefi[d_Pxy]{\Pxy, deterministic two-party communication}{
For $n\in\NN$, let the sets $\sz{\Cl A}, \sz{\Cl B}$ be such that $\Max{\sz{\Cl A}, \sz{\Cl B}}\in (2^{n-1}, 2^n]$ and let $f:\: \Cl A \times \Cl B \to \OI$.

Let $\Pi$ be a deterministic protocol where
\itstart
\item Alice receives $x$ and Bob receives $y$;
\item Alice and Bob interact;
\item Bob produces the answer.
\itend
If the transcript of $\Pi(x,y)$ contains at most $k(n)$ bits and the protocol computes $f(x,y)$, then we say that the \Pxy-complexity of $f$, denoted by $\Pxy(f)$, is at most $k(n)$.

We call a protocol efficient if its transcript contains at most $\plog(n)$ bits and we say that a function is efficiently computable in \Pxy\ if it has an efficient \Pxy-protocol.
We denote by \Pxy\ the class of total bipartite Boolean functions (or, alternatively, the languages of satisfying assignments to such functions, viewed as predicates) that are efficiently computable in \Pxy.
}

\ndefi[d_NPxy]{\NPxy, non-deterministic two-party communication}{
Let $R$ be a family of combinatorial rectangles in $\Cl A \times \Cl B$, $\sz{R}\in (2^{k(n)-1},2^{k(n)}]$.
Denote by $\Pi_R$ the corresponding \NPxy-protocol:\ it has complexity $k(n)$ and computes the predicate
\m{
f_R(x,y) \deq
\Cases
{\top}{if $(x,y)\in r$ for some $r\in R$;}
{\bot}{otherwise.}
}

For $n\in\NN$, let the sets $\sz{\Cl A}, \sz{\Cl B}$ be such that $\Max{\sz{\Cl A}, \sz{\Cl B}}\in (2^{n-1}, 2^n]$ and let $f:\: \Cl A \times \Cl B \to \TF$.
The \NPxy-complexity of $f$, denoted by $\NPxy(f)$, equals the minimal complexity of an \NPxy-protocol that computes $f(x,y)$.
We denote by \NPxy\ the class of total bipartite Boolean functions (or, alternatively, the languages of satisfying assignments to such functions, viewed as predicates) whose \NPxy-complexity is at most $\plog(n)$.
}

\sect[s_patt]{Patterned non-determinism}

While some of the definitions given next could be naturally generalised to the case of \e{partial} bipartite problems, we keep the notation simple by only considering the \e{total} case, which is of interest to us in this work.
That is, the input space will have the product structure $\Cl A \times \Cl B$.

\ndefi[d_acc_prot]{Accepting patterns of \NPxy-protocols}{
Let $\Pi$ be an \NPxy-protocol over input space $\Cl A \times \Cl B$ and let $R_\Pi$ be the set of its rectangles.

Call
\m{
\Gamma_\Pi \deq \set{\set{r\in R_\Pi}[(x,y)\in r]}[(x,y)\in\Cl A \times \Cl B]
}
the family of \pss[\Pi] \e{accepting patterns}.
}

\ndefi[d_PNPr]{\PNPrxy, rectangle-patterned \NPxy}{
For $n\in\NN$, let the sets $\sz{\Cl A}, \sz{\Cl B}$ be such that $\Max{\sz{\Cl A}, \sz{\Cl B}}\in (2^{n-1}, 2^n]$ and let $f:\: \Cl A \times \Cl B \to \TF$.

Let $\Pi$ be an \NPxy-protocol (of any cost) that computes $f$ such that the corresponding family of accepting patterns $\Gamma_\Pi$ contains at most $2^{k(n)}$ elements, then we say that the \PNPrxy-complexity of $f$, denoted by $\PNPrxy(f)$, is at most $k(n)$.

We denote by \PNPrxy\ the class of total bipartite Boolean functions (or, alternatively, the languages of satisfying assignments to such functions, viewed as predicates) whose \PNPrxy-complexity is at most $\plog(n)$.
}

Note that the above definition does not require that the \NPxy-protocol $\Pi$ used to witness the \PNPrxy-complexity of $f$ is by itself efficient.\fn[fn_drop_rec]
{
There are at most $\sz{\Gamma_\Pi}$ rectangles that are not covered by other rectangles, so by dropping ``meaningless'' rectangles recursively, any $\Pi$ can be transformed into an equivalent protocol of cost at most $\Log{\sz{\Gamma_\Pi}}$.
}

The model \PNPrxy\ is a variation of previously studied \UPxy, \FPtxy\ and \FPxy:\ they correspond to restricting \defiref{d_PNPr} by the condition that every $\gamma\in\Gamma_\Pi$ is of size at most $1$, $t$ or $\plog(n)$, respectively.
Trivially, \PNPrxy\ is a strengthening of those models (as long as $t\le\plog(n)$ in the case of \FPtxy).
On the other hand, Yannakakis~\cite{Y91_Ex_Co} proved that $\UPxy=\Pxy$ and later Karchmer, Newman, Saks and Wigderson~\cite{KNSW94_Non} strengthened it to $\FPxy=\Pxy$:\,\fn
{
Remember that all communication problems considered in this paper are total functions (for promise problems the equalities do not hold in general).
}

\nfct[f_FPxy]{\FPxy\ vs.\ \Pxy~\cite{KNSW94_Non}}{For every total Boolean $f:\: \Cl A \times \Cl B \to \TF$ and $t\in \NN$,
\m{
\Pxy(f) \in \asO{t^2\tm \FPtxy(f)^2}
.}
Accordingly, $\FPxy=\Pxy$.
}

In \sref{ss_PNPrxy} we will address the question whether $\PNPrxy=\Pxy$.

But there is a noteworthy intuitive difference between \PNPrxy\ and the classes \UPxy, \FPtxy\ and \FPxy, namely \e{the robustness with respect to interactive verification} of the corresponding definitions.
In \defiref[d_acc_prot]{d_PNPr} we treat individual rectangles of $\Pi$ as the \NPxy-witnesses, and it is natural to ask \e{what would happen to the defined models if, instead, we let the deterministic ``verifier'' be interactive}?
That is, let $\Pi'$ be an efficient deterministic protocol where Alice receives both $x\in \Cl A$ and a witness $w\in \OI^{\plog(n)}$, Bob receives $y\in \Cl B$, then they interact and either accept or reject; say that such $\Pi'$ computes the predicate $f_{\Pi'}(x_0,y_0)$ that gets the true value if and only if there exists $w_0$ such that $\Pi'((x_0,w_0),y_0)$ accepts.

Obviously, $f_{\Pi'}\in \NPxy$.
If it is additionally guaranteed that $\forall\: (x,y)\: \sz{\set{i}[f_i(x,y) = \top]} \le 1$, $\le t$ or $\le \plog(n)$, then, respectively, $f\in \UPxy$, $f\in \FPtxy$ or $f\in \FPxy$ -- trivially, as follows from the respective definitions.
The case of \PNPrxy\ is probably more interesting, as we see next.

For every $f\in \NPxy$ we will assume a disjunctive decomposition $f(x,y) = \lor_{i=1}^mf_i(x,y)$, where every $f_i$ represents the computation of the \NPxy-protocol for the fixed witness value $w=i$ -- that is, $\forall\: i:\: f_i\in \Pxy$.

\ndefi[d_acc_disj]{Accepting patterns of disjunctions}{
Let $f(x,y) = \lor_{i=1}^mf_i(x,y)$ be defined over $(x,y)\in \Cl A \times \Cl B$.

Call
\m{
\Gamma_f \deq \set{\set{i}[f_i(x,y) = \top]}[(x,y)\in\Cl A \times \Cl B]
}
the family of \pss[f] \e{accepting patterns} with respect to the decomposition $\lor_if_i(x,y)$ (often implicitly assumed).
}

\ndefi[d_PNPf]{\PNPxy, patterned \NPxy}{
For $n\in\NN$, let the sets $\sz{\Cl A}, \sz{\Cl B}$ be such that $\Max{\sz{\Cl A}, \sz{\Cl B}}\in (2^{n-1}, 2^n]$ and let $f:\: \Cl A \times \Cl B \to \TF$.

If $f$ has a decomposition
\m{
f(x,y) \g= \biglor_{i=1}^mf_i(x,y)
,}
such that the corresponding family of accepting patterns $\Gamma_f\sbseq \pow{[m]}$ contains at most $2^{k(n)}$ elements and $\forall\: i:\: \Pxy(f_i)\le k(n)$, then we say that the \PNPxy-complexity of $f$, denoted by $\PNPxy(f)$, is at most $k(n)$.\,\fn[fn_drop_fi]
{
It is not required by the definition, but can be assumed without loss of generality that $m\le \sz{\Gamma_f}$:\ the set $\set{i_0\in [m]}[\exists\: (x,y)\in \Cl A\times \Cl B:\: f_{i_0}(x,y)=\top,\, \lor_{j\ne i_0}f_j(x,y)=\bot]$ contains at most $\sz{\Gamma_f}$ elements and all other \pl[f_i] are ``meaningless'' and can be recursively dropped from the decomposition $f(x,y) = \lor_if_i(x,y)$ without affecting $f(x,y)$ (cf.~Footnote~\ref{fn_drop_rec}).
}

We denote by \PNPxy\ the class of total bipartite Boolean functions (or, alternatively, the languages of satisfying assignments to such functions, viewed as predicates) whose \PNPxy-complexity is at most $\plog(n)$.
}

Obviously, $\PNPrxy \sbseq \PNPxy$.
The question whether the two complexity classes are equal will require our further attention:
In particular, the assumption $[\sz{\Gamma_f}\le 2^{k(n)}]$ doesn't have immediate implications regarding the number of possible accepting sets of rectangles in the (assumed) \Pxy-protocols for \pl[f_i]; what is more, there doesn't have to exist an efficient witness that $[\set{i}[f_i(x,y) = \top]=s]$ as long as $\sz{s}$ is large (say, $n^{\asOm1}$) -- in contrast to the case of \PNPrxy, where the corresponding witness would be the intersection of $\sz{s}$ rectangles, thus itself a rectangle.
See \sref{ss_PNPxy} (\lemref{l_PNPxy} in particular).

\ssect[ss_PNPrxy]{Rectangle-patterned non-determinism (\PNPrxy) vs.\ determinism (\Pxy)}[Rectangle-patterned non-determinism vs.\ determinism]

Are the complexity classes \PNPrxy\ and \Pxy\ equal?

\lem[l_PNPrxy]{For every total Boolean $f:\: \Cl A \times \Cl B \to \TF$,
\m{
\Pxy(f) \in \asO{\PNPrxy(f)^2}
.}
Accordingly, $\PNPrxy=\Pxy$.
}

\prfstart
Let $\Pi_f$ be an \NPxy-protocol that computes $f$ and witnesses (cf.~\defiref{d_PNPr}) that $\PNPrxy(f)\le k(n)$.
Let $R_\Pi$ be the set of \pss[\Pi_f] rectangles and $\Gamma_\Pi\sbseq \pow{R_\Pi}$ be the corresponding family of accepting patterns ($\sz{\Gamma_\Pi}\le 2^{k(n)}$).

Consider the following \Pxy-protocol $\Phi$ for input $(x,y)\in \Cl A\times \Cl B$:

\begin{mdframed}
\enstart
\item
$j\asgn0$; $\Cel A_1\asgn\Cl A$; $\Cel B_1\asgn\Cl B$.
\item\label{step_LoopBeg}
\itstart\setlength{\itemindent}{-1em}
\item $j\asgn j+1$;
\item
\f{
\Gamma_j\asgn \set{\set{r\in R_\Pi}[(x',y')\in r]}[(x',y')\in\Cel A_j\times \Cel B_j]
.}
\itend
\item\label{step_Check1}
If there exists $r_A\times r_B=r\in R_\Pi$ such that
\m{
\sz{\set{\gamma\in \Gamma_j}[r\in\gamma]}
\ge \fr 13\tm \sz{\Gamma_j}
>0
,}
then do:
\itstart
\item if $(x,y)\in r$, then output ``$\top$'' and \ebi{halt};
\item if $x\nin r_A$, then let $\Cel A_{j+1}\asgn \Cel A_j\smin r_A$, else $\Cel A_{j+1}\asgn \Cel A_j$;
\item if $y\nin r_B$, then let $\Cel B_{j+1}\asgn \Cel B_j\smin r_B$, else $\Cel B_{j+1}\asgn \Cel B_j$;
\item go to Step~\ref{step_LoopBeg}.
\itend
\item\label{step_Check2a}
If there exists $r_A\times r_B=r\in R_\Pi$ such that $x\in r_A$ and
\m{
\sz{\set{\gamma\in \Gamma_j}[\forall\: x'\in \Cel A_j\cap r_A:\: \exists\: r_A'\times r_B'\in \gamma:\: x'\nin r_A']}
\ge \fr 13\tm \sz{\Gamma_j}
>0
,}
then do:
\itstart
\item if $y\in r_B$, then output ``$\top$'' and \ebi{halt};
\item $\Cel A_{j+1}\asgn \Cel A_j\cap r_A$;
\item $\Cel B_{j+1}\asgn \Cel B_j$;
\item go to Step~\ref{step_LoopBeg}.
\itend
\item\label{step_Check2b}
If there exists $r_A\times r_B=r\in R_\Pi$ such that $y\in r_B$ and
\m{
\sz{\set{\gamma\in \Gamma_j}[\forall\: y'\in \Cel B_j\cap r_B:\: \exists\: r_A'\times r_B'\in \gamma:\: y'\nin r_B']}
\ge \fr 13\tm \sz{\Gamma_j}
>0
,}
then do:
\itstart
\item if $x\in r_A$, then output ``$\top$'' and \ebi{halt};
\item $\Cel A_{j+1}\asgn \Cel A_j$;
\item $\Cel B_{j+1}\asgn \Cel B_j\cap r_B$;
\item go to Step~\ref{step_LoopBeg}.
\itend
\item\label{step_End}
Output ``$\bot$'' and \ebi{halt}.
\enend
\end{mdframed}

We claim that $\Phi(x,y)$ has complexity \asO{k(n)^2} and computes $f(x,y)$.

\para[par_AB_as]{At the end of Step~\ref{step_LoopBeg}
\m{
(x,y)\in \Cel A_j\times \Cel B_j
}
always.}
In the beginning this is trivially true, and the updates (shrinkages) of $\Cel A$ and $\Cel B$ in Steps~\ref{step_Check1}, \ref{step_Check2a} and \ref{step_Check2b} occur under conditions that guarantee that $(x,y)$ stays inside $\Cel A_{j+1}\times \Cel B_{j+1}$.

\para[par_Gj_as]{At the end of Step~\ref{step_LoopBeg}
\m{
\Gamma_j = \set{\set{r\in R_\Pi}[(x',y')\in r]}[(x',y')\in\Cel A_j\times \Cel B_j]
}
always.}

\para{Answer ``$\top$'' is always correct.}
Indeed, producing such an answer necessarily represents having found $r\in R_\Pi$ such that $(x,y)\in r$, therefore $f(x,y)=\top$.

\para{Answer ``$\bot$'' is always correct.}
Let $j_0$ be the value of the index $j$ when ``$\bot$'' has been produced at Step~\ref{step_End} and assume towards contradiction that $f(x,y)=\top$ and therefore $(x,y)\in r_A\times r_B\in R_\Pi$.
If $\sz{\Gamma_{j_0}}=0$, then the desired contradiction follows readily from \paref[par_AB_as]{par_Gj_as}, so assume that $\sz{\Gamma_{j_0}}>0$.

As the entry condition of Step~\ref{step_Check1} was unsatisfied, it must be the case that
\m{
\sz{\set{\gamma\in \Gamma_{j_0}}[r_A\times r_B\in\gamma]}
< \fr 13\tm \sz{\Gamma_{j_0}}
,}
and therefore $|\wtl{\Gamma_{j_0}}| > \dr23\tm |\Gamma_{j_0}|$ for
\m{
\wtl{\Gamma_{j_0}} \deq
\set{\gamma\in \Gamma_{j_0}}[r_A\times r_B\nin\gamma]
.}
Due to \paref{par_Gj_as}, $\forall\: \gamma_0\in\wtl{\Gamma_{j_0}}$, $\forall\: (x',y')\in\Cel A_{j_0}\times \Cel B_{j_0}$:
\m{
\lf(\forall\ r'\in \gamma_0:\: (x',y')\in r'\rt)
~\Then~ (x',y')\nin r_A\times r_B
,}
that is,
\m{
(x',y')\in r_A\times r_B
~\Then~ \exists\: r_A'\times r_B'\in\gamma_0:\:
x'\nin r_A' \:\lor\: y'\nin r_B'
.}

As $(x',y')$ can be any pair from the product set $\Cel A_{j_0}\times \Cel B_{j_0}$,
the above readily decomposes into
\m{
\forall\: x'\in \Cel A_{j_0}\cap r_A:\: \exists\: r_A'\times r_B'\in \gamma_0:\: x'\nin r_A'
~\biglor~
\forall\: y'\in \Cel B_{j_0}\cap r_B:\: \exists\: r_A'\times r_B'\in \gamma_0:\: y'\nin r_B'
.}
In other words, $\wtl{\Gamma_{j_0}} = \wtl{\Gamma_{j_0}^A} \cup \wtl{\Gamma_{j_0}^B}$ (not necessarily disjointly), where
\m{
\wtl{\Gamma_{j_0}^A} \deq
\set{\gamma\in \Gamma_{j_0}}[\forall\: x'\in \Cel A_{j_0}\cap r_A:\: \exists\: r_A'\times r_B'\in \gamma:\: x'\nin r_A']
}
and
\m{
\wtl{\Gamma_{j_0}^B} \deq
\set{\gamma\in \Gamma_{j_0}}[\forall\: y'\in \Cel B_{j_0}\cap r_B:\: \exists\: r_A'\times r_B'\in \gamma:\: y'\nin r_B']
.}
As $|\wtl{\Gamma_{j_0}}| > \dr23\tm |\Gamma_{j_0}|$, it necessarily holds that $|\wtl{\Gamma_{j_0}^A}| > \dr13\tm |\Gamma_{j_0}|$ or $|\wtl{\Gamma_{j_0}^B}| > \dr13\tm |\Gamma_{j_0}|$, and therefore the entry condition of Step~\ref{step_Check2a} or \ref{step_Check2b} must have been satisfied, contradicting our assumption that ``$\bot$'' was produced at Step~\ref{step_End}.

\para{The protocol makes \asO{k(n)} iterations.}
Due to \paref{par_AB_as}, it is guaranteed by the entry conditions and the actions of Steps~\ref{step_Check1}, \ref{step_Check2a} and \ref{step_Check2b} that
\m{
\sz{\Gamma_j} \le \sz{\Gamma_{j-1}} \tm \dr23
}
at every protocol round $j>1$.
And we have assumed that $\sz{\Gamma_\Pi}\le 2^{k(n)}$.

\para{The \Pxy-complexity of one iteration of the protocol is in \asO{k(n)}.}
As the protocol proceeds, both players locally keep track of $j$, $\Gamma_j$, $\Cel A_j$ and $\Cel B_j$.
Non-trivial are only Steps~\ref{step_Check1}, \ref{step_Check2a} and \ref{step_Check2b}, and it is easy to see that their entry conditions can be checked locally by at least one of the players, and the actions (basically, checking whether $(x,y)\in r=r_A\times r_B$) require \asO{\Log{|R_\Pi|}} bits of communication (the cost of sending a ``pointer'' to $r\in R_\Pi$), which can be assumed to be in \asO{k(n)} (cf.~Footnote~\ref{fn_drop_rec}).
\prfend[\lemref{l_PNPrxy}]

\ssect[ss_PNPxy]{Patterned non-determinism (\PNPxy) vs.\ determinism (\Pxy)}[Patterned non-determinism vs.\ determinism]

As mentioned earlier, the communication model \PNPxy\ is an interesting object of study because, in particular, the transition from its ``rectangular'' version \PNPrxy\ to the general case looks challenging.\fn
{
E.g., if we look at the protocol $\Phi(x,y)$ from \sref{ss_PNPrxy}, then, first of all, it is not clear how to efficiently generalise for the case of \PNPxy\ the entry conditions of Steps~\ref{step_Check2a} and \ref{step_Check2b}; what is more, the logic underlying those conditions (as represented by the analysis of $\Phi$) doesn't seem to generalise readily.
}

\lem[l_PNPxy]{For every total Boolean $f:\: \Cl A \times \Cl B \to \TF$,
\m{
\Pxy(f) \in \asO{\PNPxy(f)^6}
.}
Accordingly, $\PNPxy=\PNPrxy=\Pxy$.
}

To prove it we will use the following simple ``hitting set'' statement.

\clm[c_s_good]{Let $\Gamma\sbseq \pow{[m]}$ and $t\in \NN$ be such that
\m{
\forall\: \gamma\in \Gamma:\: \sz{\gamma}\le 2t
.}
Then there exists $\sigma\sbseq [m]$ such that
\m{
\Max[\gamma\in \Gamma]{\sz{\gamma\cap\sigma}}
\le 8e+\log\sz{\Gamma}
}
and
\m{
\sz{\set{\gamma\in \Gamma}[\sz{\gamma}\ge t,\, \gamma\cap\sigma \ne\es]}
\ge \fr 12\tm \sz{\set[:]{\gamma\in \Gamma}[\sz{\gamma}\ge t]}
.}
}

\prfstart
Let $\sigma_0\unin\chs{[m]}{\dr {2m}t}$, then
\m{
\forall\: \gamma\in \Gamma, \sz{\gamma}\ge t:\:
\PR[\sigma_0]{\gamma\cap\sigma_0=\es}
\le \lf(\fr{m-t}{m}\rt)^{\dr {2m}t}
= \lf(1-\fr tm\rt)^{\dr {2m}t}
< \fr14
}
and
\m{
\E[\sigma_0]{\sz{\set[:]{\gamma\in \Gamma}[\sz{\gamma}\ge t,\, \gamma\cap\sigma_0 =\es]}}
< \fr14\tm \sz{\set[:]{\gamma\in \Gamma}[\sz{\gamma}\ge t]}
,}
so
\m[m_Pr_keep]{
\PR[\sigma_0]
{\sz{\set[:]{\gamma\in \Gamma}[\sz{\gamma}\ge t,\, \gamma\cap\sigma_0 \ne\es]}
\ge \fr 12\tm \sz{\set[:]{\gamma\in \Gamma}[\sz{\gamma}\ge t]}}
> \fr12
.}

Denote $s\deq 8e+\log\sz{\Gamma}$.
As $\forall\: \gamma\in \Gamma:\: \sz{\gamma}\le 2t$,
\m{
\forall\: \gamma\in \Gamma:\:
\PR[\sigma_0]{\sz{\gamma\cap\sigma_0}\ge s}
\le \chs {2t}s\tm \lf(\fr{\dr {2m}t}{m}\rt)^s
\le \lf(\fr{2et}{s}\tm \fr 2t\rt)^s
=\lf(\fr{4e}s\rt)^s
\le \fr{2^{-8e}}{\sz{\Gamma}}
,}
that is
\m{
\PR[\sigma_0]{\exists\: \gamma\in \Gamma:\: \sz{\gamma\cap\sigma_0}\ge s}
\le 2^{-8e}
.}
Together with~\bref{m_Pr_keep} this implies the result.
\prfend[\clmref{c_s_good}]

\prfstart[\lemref{l_PNPxy}]

Let
\m{
f(x,y) \g= \biglor_{i=1}^mf_i(x,y)
}
be a decomposition that witnesses (cf.~\defiref{d_PNPf}) that $\PNPxy(f)\le k(n)$.
Let $\Gamma_f\sbseq \pow{[m]}$ be the corresponding family of accepting patterns ($\sz{\Gamma_f}\le 2^{k(n)}$).
Assume without loss of generality that $m\le \sz{\Gamma_f}$ (cf.~Footnote~\ref{fn_drop_fi}).
For every $i\in [m]$, let $\Pi_i$ be a \Pxy-protocol of complexity at most $k(n)$ that computes $f_i(x,y)$ and let $R_i$ be the set of \pss[\Pi_i] \e{accepting} rectangles ($\sz{R_i}\le 2^{k(n)}$).
Denote for any non-empty $s\sbseq[m]$:
\m{
R_s \deq \set{r_1\ds[\cap] r_{|s|}}[r_l\in R_{s(l)} \txt{ for } 1\le l\le |s|]
,}
that is, $R_s$ is the family of rectangle intersections -- therefore rectangles themselves -- that witness $[\bigland_{i\in s}f_i(x,y)]$ for $(x,y)\in \Cl A\times \Cl B$.
Clearly, $\sz{R_s} \le 2^{|s|\tm k(n)}$.

Consider the following \Pxy-protocol $\Psi$ for input $(x,y)\in \Cl A\times \Cl B$:

\begin{mdframed}
\enstart
\item
$j\asgn0$; $\Cel A_1\asgn\Cl A$; $\Cel B_1\asgn\Cl B$;
$s_0\asgn\Max[\gamma\in \Gamma_f]{\sz{\gamma}}$; $\sigma_0\asgn\es$.
\item\label{step_LoopBeg_Psi} 
\itstart\setlength{\itemindent}{-1em}
\item $j\asgn j+1$;
\item
\f{
\Gamma_j\asgn \set{\set{i}[f_i(x',y') = \top]}[(x',y')\in\Cel A_j \times \Cel B_j]
.}
\itend
\item\label{step_End_Psi}
If $\Gamma_j =\set{\es}$, then output ``$\bot$'' and \ebi{halt}.
\item \label{step_sigma_Psi}
If $\set{\gamma\cap\sigma_{j-1}}[\gamma\in \Gamma_j] = \set{\es}$, then do:
\itstart
\item
if $\Gamma_j\cap[\dr{s_{j-1}}2,\, s_{j-1}]=\es$, then let $s_j\asgn\Max[\gamma\in \Gamma_j]{\sz{\gamma}}$, else $s_j\asgn s_{j-1}$;
\item
let $\sigma_j\sbseq [m]$ be (as guaranteed by \clmref{c_s_good}) such that
\m{
\sz{\set{\gamma\in \Gamma_j\cap[\dr{s_j}2,\, s_j]}[\gamma\cap\sigma_j \ne\es]}
\ge \fr 12\tm \sz{\Gamma_j\cap[\dr{s_j}2,\, s_j]}
}
and
\m{
\Max[\gamma\in \Gamma_j]{\sz{\gamma\cap\sigma_j}}
\le 8e+\log\sz{\Gamma_j}
;}
\itend
else:
\itstart
\item $s_j\asgn s_{j-1}$;
\item $\sigma_j\asgn\sigma_{j-1}$.
\itend
\item \label{step_ass_Psi}
\itstart\setlength{\itemindent}{-1em}
\item $t_j\asgn\Max[\gamma\in \Gamma_j]{\sz{\gamma\cap\sigma_j}}$;
\item $\Delta_j\asgn\set[:]{\gamma\cap\sigma_j}[\gamma\in \Gamma_j,\, \sz{\gamma\cap\sigma_j} = t_j]$;
\item
\f{
\Pi_j\asgn
\set{\Cel A_j \times \Cel B_j\cap r}
[r\in R_\delta \txt{ for } \delta\in \Delta_j] \smin \set{\es}
;}
\item 
\f{
\Pi_j^A\asgn
\set[:]{r_A\times r_B\in \Pi_j}[\sz{\set[:]{r_A'\times r_B'\in \Pi_j}
[r_A\cap r_A'=\es]} \ge \fr{\sz{\Pi_j}-1}2]
;}
\item 
\f{
\Pi_j^B\asgn
\set[:]{r_A\times r_B\in \Pi_j}[\sz{\set[:]{r_A'\times r_B'\in \Pi_j}
[r_B\cap r_B'=\es]} \ge \fr{\sz{\Pi_j}-1}2]
.}
\itend
\item \label{step_Check1_Psi}
If there exists $r_A\times r_B\in \Pi_j^A$ such that $x\in r_A$, then do:
\itstart
\item if $y\in r_B$, then output ``$\top$'' and \ebi{halt};
\item $\Cel A_{j+1}\asgn \Cel A_j\cap r_A$;
\item $\Cel B_{j+1}\asgn \Cel B_j\smin r_B$;
\item go to Step~\ref{step_LoopBeg_Psi}.
\itend
\item \label{step_Check2_Psi}
If there exists $r_A\times r_B\in \Pi_j^B$ such that $y\in r_B$, then do:
\itstart
\item if $x\in r_A$, then output ``$\top$'' and \ebi{halt};
\item $\Cel A_{j+1}\asgn \Cel A_j\smin r_A$;
\item $\Cel B_{j+1}\asgn \Cel B_j\cap r_B$;
\item go to Step~\ref{step_LoopBeg_Psi}.
\itend
\item \label{step_Else_Psi}
\itstart\setlength{\itemindent}{-1em}
\item $\Cel A_{j+1}\asgn \Cel A_j\smin \bigcup_{r_A\times r_B\in \Pi_j^A}r_A$;
\item $\Cel B_{j+1}\asgn \Cel B_j\smin \bigcup_{r_A\times r_B\in \Pi_j^B}r_B$;
\item go to Step~\ref{step_LoopBeg_Psi}.
\itend
\enend
\end{mdframed}

We claim that $\Psi(x,y)$ has complexity \asO{k(n)^6} and computes $f(x,y)$.

In the following analysis we call a \f j-indexed value (e.g., $s_j$ or $\sigma_j$) \e{unchanged} as long as the next round's value is the same as the last round's (e.g., due to assignments like $s_j\asgn s_{j-1}$ or $\sigma_j\asgn\sigma_{j-1}$).
Otherwise we will say that the corresponding value \e{changes} at round $j$.

\para[par_AB_as_Psi]{At the end of Step~\ref{step_LoopBeg_Psi}
\m{
(x,y)\in \Cel A_j\times \Cel B_j
}
always.}
In the beginning this is trivially true.
The updates (shrinkages) of $\Cel A$ and $\Cel B$ in Steps~\ref{step_Check1_Psi} and \ref{step_Check2_Psi} occur under conditions that guarantee that $(x,y)$ stays inside $\Cel A_{j+1}\times \Cel B_{j+1}$.
The updates in Step~\ref{step_Else_Psi} occur only if the entry conditions of both Steps~\ref{step_Check1_Psi} and \ref{step_Check2_Psi} were unsatisfied, which also guarantees that $(x,y)$ stays inside.

\para[par_Gj_as_Psi]{At the end of Step~\ref{step_LoopBeg_Psi}
\m{
\Gamma_j = \set{\set{i}[f_i(x',y') = \top]}[(x',y')\in\Cel A_j \times \Cel B_j]
}
always.}

\para[par_Psi_top]{Answer ``$\top$'' is always correct.}
Indeed, producing such an answer necessarily represents having found $r\in R_s$ for some non-empty $s\sbseq[m]$ such that $(x,y)\in r$ -- that is, the input pair is inside a non-empty intersection of \pss[\Pi_i] accepting rectangles, so $f(x,y)=f_i(x,y)=\top$.

\para{Answer ``$\bot$'' is always correct.}
Answering ``$\bot$'' in Step~\ref{step_End_Psi} is conditioned upon $[\Gamma_j =\set{\es}]$, due to \paref[par_AB_as_Psi]{par_Gj_as_Psi} this implies that $f(x,y)=\bot$.

\para[par_s_j_Psi]{The value of $s_j$ changes at most $\log\sz{\Gamma_f}$ times.}
This can only happen in Step~\ref{step_sigma_Psi} if the condition $\lf[\Gamma_j\cap[\dr{s_{j-1}}2,\, s_{j-1}]=\es\rt]$ is satisfied.
We have $s_0=\Max[\gamma\in \Gamma_f]{\sz{\gamma}}\le m\le \sz{\Gamma_f}$ and every time $s_j$ changes, it is necessarily the case both that $s_j< \dr{s_{j-1}}2$ and $s_j>0$ (the latter is due to the check in Step~\ref{step_End_Psi}).

\para[par_sigma_Psi]{While $s_j$ remains unchanged, $\sigma_j$ changes at most $\log\sz{\Gamma_f}+1$ times.}
The change can happen only in Step~\ref{step_sigma_Psi} if the condition $\lf[\set{\gamma\cap\sigma_{j-1}}[\gamma\in \Gamma_j] = \set{\es}\rt]$ is satisfied.
As long as the value of $s_j$ remains unchanged, every redefinition of $\sigma_j$ results in $\set{\gamma\cap\sigma_j}[\gamma\in \Gamma_j]$ containing at least half of $\Gamma_j\cap[\dr{s_{j-1}}2,\, s_{j-1}]$, and if $\Gamma_j$ changes, then its content necessarily shrinks -- accordingly, there can be at most $\log\sz{\Gamma_j}+1 \le \log\sz{\Gamma_f}+1$ redefinitions of $\sigma_j$ for the same value of $s_j$.

\para[par_t_j_Psi]{While $\sigma_j$ and $s_j$ remain unchanged, the value of $t_j$ either remains unchanged or decreases (Step~\ref{step_ass_Psi}); $t_j\le 8e+\log\sz{\Gamma_f}$ always.
}

\para[par_iter_in_Psi]{While $\sigma_j$, $s_j$ and $t_j$ remain unchanged, the protocol makes \asO{k(n)^2} iterations.}
Intuitively, in this situation our protocol solves with respect to $(x,y)\in\Cel A_j\times \Cel B_j$ the problem
\m{
\txt{accept if }
\sz{\set{i\in \sigma_j}[f_i(x,y)=\top]} = t_j
,}
while it is guaranteed by definition that
\m[m_le_t_j]{
\forall\: (x',y')\in\Cel A_j\times \Cel B_j:\:
\sz{\set{i\in \sigma_j}[f_i(x',y')=\top]} \le t_j
.}
Then $\Delta_j$ is the set of accepting patterns (in the sense analogous to \defiref{d_acc_disj}, but with ``$\biglor$'' replaced by $t_j$-threshold) and $\Pi_j$ is the corresponding family of witnessing rectangle intersections, therefore rectangles themselves.
As the updates only can shrink the sets $\Cel A_j$ and $\Cel B_j$, also the family $\Pi_j$ only shrinks while $\sigma_j$, $s_j$ and $t_j$ remain unchanged.

We claim that $\Pi_j = \Pi_j^A\cup\Pi_j^B$ (not necessarily disjointly).
Towards contradiction, assume the opposite and let $r_A\times r_B\in \Pi_j\smin \Pi_j^A\smin \Pi_j^B$, then there exists $r_A'\times r_B'\in \Pi_j$ such that $r_A'\times r_B'\ne r_A\times r_B$ but $r_A'\times r_B'\cap r_A\times r_B\ne\es$.
Let $(x_0,y_0)\in r_A'\times r_B'\cap r_A\times r_B$, $r_A\times r_B=\Cel A_j \times \Cel B_j\cap r$ and $r_A'\times r_B'=\Cel A_j \times \Cel B_j\cap r'$ for $r\in R_\delta$, $r'\in R_{\delta'}$ and $\delta,\, \delta'\in \Delta_j$.
If $\delta = \delta'$, then $r$ and $r'$ are distinct elements of
\m{
R_\delta = \set{r_1\ds[\cap] r_{|s|}}
[r_l\in R_{\delta(l)} \txt{ for } 1\le l\le |\delta|]
,}
contradicting the assumption that each $R_i$ is the set of accepting rectangles in a \e{deterministic} protocol (whose rectangles are therefore disjoint).
If, on the other hand, $\delta \ne \delta'$, then $\delta \cup \delta'$ is a subset of
\m{
\set{i\in \sigma_j}[f_i(x_0,y_0)=\top]
,}
contradicting~\bref{m_le_t_j}, as $\sz{\delta \cup \delta'} > t_j$.
So,
\m[m_Pi_j_A_B]{
\Pi_j = \Pi_j^A\cup\Pi_j^B
.}

Now assume that at round $j+1$ the values of $\sigma_{j+1}$, $s_{j+1}$ and $t_{j+1}$ remain unchanged.
There are cases to consider.

If the instruction ``go to Step~\ref{step_LoopBeg_Psi}'' has been performed at Step~\ref{step_Check1_Psi} of round $j$ (the case of Step~\ref{step_Check2_Psi} is similar), then $x\in r_A$ such that
\m{
\sz{\set[:]{r_A'\times r_B'\in \Pi_j}
[r_A\cap r_A'=\es]} \ge \fr{\sz{\Pi_j}-1}2
.}
As
$\Pi_{j+1} = \set{\Cel A_{j+1} \times \Cel B_{j+1}\cap r}
[r\in R_\delta,\, \delta\in \Delta_{j+1}] \smin \set{\es}$,
the assignment $\Cel A_{j+1}\asgn \Cel A_j\cap r_A$ at Step~\ref{step_Check1_Psi} of round $j$ has ``removed'' at least $\dr{(\sz{\Pi_j}-1)}2$ elements from $\Pi_{j+1}$ in comparison to $\Pi_j$, and the assignment $\Cel B_{j+1}\asgn \Cel B_j\smin r_B$ has removed at least one more (as $r_A\times r_B\in \Pi_j$).
Overall, $\sz{\Pi_{j+1}} \le \dr{\sz{\Pi_j}}2$.

If, on the other hand, ``go to Step~\ref{step_LoopBeg_Psi}'' has been performed at Step~\ref{step_Else_Psi} of round $j$, then the preceding assignments $\Cel A_{j+1}\asgn \Cel A_j\smin \bigcup_{r_A\times r_B\in \Pi_j^A}r_A$ and $\Cel B_{j+1}\asgn \Cel B_j\smin \bigcup_{r_A\times r_B\in \Pi_j^B}r_B$ guarantee -- along with \bref{m_Pi_j_A_B} -- that $\sz{\Pi_{j+1}} =\es$ (which, in fact, contradicts our assumption that $\sigma_j$, $s_j$ and $t_j$ remain unchanged at round $j+1$).

Accordingly, while $\sigma_j$, $s_j$ and $t_j$ remain unchanged since round $j_0$, the protocol can make only \asO{\log\sz{\Pi_{j_0}}} iterations. 
Since $\Delta_{j_0}\sbseq \chs{[m]}{t_{j_0}}$ and $\sz{\Pi_{j_0}} \le \sum_{\delta\in \Delta_{j_0}} \sz{R_\delta}$
by definition, it holds -- as required -- that
\m{
\sz{\Pi_{j_0}} \le 2^{t_{j_0}\tm k(n)} \tm \sz{\Delta_{j_0}}
\le 2^{\asO{k(n)^2}}
,}
as $t_{j_0}\le 8e+\log\sz{\Gamma_{j_0}}$ and $\sz{\Delta_{j_0}} \le \sz{\Gamma_{j_0}} \le \sz{\Gamma_f}\le 2^{k(n)}$.

\para{The protocol makes \asO{k(n)^5} iterations.}
Follows readily from \paref[par_s_j_Psi][par_sigma_Psi][par_t_j_Psi]{par_iter_in_Psi}.

\para{The \Pxy-complexity of one iteration of the protocol is in \asO{k(n)}.}
To agree upon the value of $\sigma_j$ in Step~\ref{step_sigma_Psi}, the players can, for instance, always pick the lexicographically first suitable candidate (which can be done locally as long as $\Gamma_j$ and $s_j$ are known to both players).
The rest is very similar to the case of $\Phi(x,y)$ in the proof of \lemref{l_PNPrxy}.
\prfend

\ssect[ss_Search]{Efficient exhaustive witness-searching in \PNPxy}[Efficient exhaustive witness-searching in PNP]

Assume that certain communication complexity subclass $\Cel C\sbseq \NPxy$ is inside \Pxy\ for the case of total functions (but not, in general, for the partial-functions case):\ as discussed earlier, some examples of such subclasses are $\NPxy \cap \coNPxy$, \UPxy, \FPxy\ and \PNPxy.
Let $f:\: \Cl A \times \Cl B \to \TF$ belong to $\Cel C$, does it necessarily follow that \NPxy-witnesses for every $(x,y)\in f^{-1}(\top)$ can be efficiently found?

The answer depends on the precise notion of witness that we have in mind.
In particular, as $f\in \Cel C\sbseq \Pxy$, there is an efficient deterministic protocol that computes $f$ and the transcript of that protocol on any input $(x_0,y_0)$ ``witnesses'' the value of $f(x_0,y_0)$.

On the other hand, we may consider a specific ``canonical'' \NPxy-protocol $\Pi'$ for $f$ that witnesses the membership $f\in \Cel C$ (recall that $\Cel C$ is a subclass of \NPxy) -- is it necessarily the case that finding a valid $\Pi'$-witness for every $(x,y)\in f^{-1}(\top)$ can be done efficiently?

It is so indeed for the cases of \UPxy\ and \FPxy:\ if $\Pi'$ computes $f$ with at most $\plog(n)$ distinct witnesses, then a $\Pi'$-witness for every $(x,y)\in f^{-1}(\top)$ can be found via essentially the same \Pxy-protocol that is used in the proof of ``$\FPxy = \Pxy$''.

It is very similar for \PNPxy:

\crl[crl_PNP_i]{
Let $f:\: \Cl A \times \Cl B \to \TF$ be of \PNPxy-complexity at most $k(n)$, as witnessed via the decomposition $f(x,y) = \lor_{i=1}^mf_i(x,y)$.

Then there exists a deterministic protocol of cost \asO{k(n)^6} that receives an input pair $(x_0,y_0)\in \Cl A \times \Cl B$ such that $f(x_0,y_0) = \top$ and outputs some $i_0$ such that $f_{i_0}(x_0,y_0) = \top$.
}

\prfstart
The protocol $\Psi$ from the proof of \lemref{l_PNPxy} finds such $i_0$ whenever it outputs ``$\top$'' (cf.~\paref{par_Psi_top} of that proof).
\prfend[\crlref{crl_PNP_i}]

The situation is different in the case of $\Cel C = \NPxy \cap \coNPxy$ (it was shown by Aho, Ullman and Yannakakis~\cite{AUY83_On} that the class was equal to \Pxy).
In~\cite{G20_The} subsets $\Cl A, \Cl B \sbseq \chs{[n]}{n^{\dr35}}$ were presented such that $\forall x\in \Cl A, y\in \Cl B:\: x\cap y\ne \es$, but finding an element from the intersection in a uniformly-random pair $(x,y)\in \Cl A\times \Cl B$ required a randomised communication protocol of complexity \asOm{\sq[5]n}.
If we define $\Cl A'\deq \Cl A\cup\set\es$, this will result in a (somewhat) non-trivial instance of the \e{set intersection problem} with respect to $(x,y)\in \Cl A'\times \Cl B$.
We can apply the following reasoning:
\itstart
\item the problem is in $\NPxy$:\ denote by $\Pi$ the \NPxy-protocol that accepts $(x,y)$ if and only if it receives some $i\in x\cap y$ as a witness;
\item the problem is in $\coNPxy$, as $x\cap y=\es$ only happens when $x=\es$ and this condition can be easily checked even without a witness;
\item due to~\cite{AUY83_On}, the corresponding set intersection problem is in $\NPxy \cap \coNPxy = \Pxy$;
\item nevertheless, given an input pair $(x,y)$ such that $x\cap y\ne \es$, finding a valid witness for $\Pi(x,y)$ cannot be done efficiently (even by a randomised protocol).
\itend

Can we strengthen \lemref{l_PNPxy} and \crlref{crl_PNP_i} even further?
Namely, if the membership $f \in \PNPxy$ is established via considering the decomposition $f(x,y) = \lor_{i=1}^mf_i(x,y)$, can we efficiently find for every given input $(x,y)\in f^{-1}(\top)$ the \e{exhaustive} list of the corresponding \NPxy-witnesses, that is, the exact content of $\set{i}[f_i(x,y) = \top]$?

Note that a positive answer wouldn't follow trivially from the repeated application of the efficient witness-finding protocol of \crlref{crl_PNP_i}:\ unlike \FPxy, \PNPxy\ allows arbitrarily large sets of witnesses for the same $(x,y)\in f^{-1}(\top)$, and therefore such repeated application until all valid witnesses are exhausted can be inefficient.
On the other hand, efficiency considerations do not readily lead to the negative answer either:\ $\Pi$ is a \PNPxy-protocol and therefore it admits at most $2^{\plog(n)}$ different \e{patters} (that is, possible exhaustive sets of valid witnesses) -- accordingly, a deterministic protocol of complexity $\plog(n)$ can have enough distinct ``leaves'' for returning every answer at least once (each protocol leaf is marked by the corresponding answer and every possible pattern must be the answer corresponding to at least one leaf).

As $f$ itself is less relevant for the problem of \e{exhaustive} witness-searching, we are switching to the functions $f_i$ as our primary objects of concern.

\ntheo[t_PNP_search]{Efficient exhaustive \PNPxy-search}{
Let $f_1(x,y)\dc f_m(x,y):\: \Cl A \times \Cl B \to \TF$ be such that $\forall\: i:\: \Pxy(f_i)\le k(n)$ and
\m{
\sz{
\set[:]{\set{i\in [m]}[f_i(x,y)=\top]}[(x,y)\in \Cl A \times \Cl B]
} \le 2^{\ell(n)}
.\,\fnm
}
\fnt
{
Here $m$ can be any function of $n$, cf.~Footnotes~\ref{fn_drop_rec} and~\ref{fn_drop_fi}.
}

Then there exists a deterministic protocol of cost \asO{k(n)^6\tm\ell(n)} that receives an input pair $(x_0,y_0)\in \Cl A \times \Cl B$ and outputs the set $\set{i}[f_i(x_0,y_0) = \top]$.
}

That is, the \e{exhaustive-search function} $F(x,y)\deq \set{i}[f_i(x,y)=\top]$ is in \Pxy\ if $f(x,y) \g= \lor_{i=1}^mf_i(x,y)$ is a bipartite total function in \PNPxy.
The statement is optimal from the structural perspective:\ if \pl[f_i] were not in \Pxy, then $F$ would not be there either; on the other hand, any deterministic protocol for $F$ must have at least $\sz{\set{\set{i}[f_i(x,y)=\top]}_{x,y}}$ leaves (as discussed above).

\prfstart
Consider the following \Pxy-protocol $\Xi$ for input $(x,y)\in \Cl A\times \Cl B$:

\begin{mdframed}
\enstart
\item
$j\asgn0$; $\Gamma_1\asgn \set[:]{\set{i\in [m]}[f_i(x,y)=\top]}[(x,y)\in \Cl A \times \Cl B]$.
\item\label{step_LoopBeg_Xi}
\itstart\setlength{\itemindent}{-1em}
\item $j\asgn j+1$;
\item
\f{
\Cel W_j^+ \asgn \set{i\in [m]}[{\sz{\set{\gamma\in \Gamma_j}[i\in \gamma]}\ge \dr{\sz{\Gamma_j}}2}]
;}
\item $\Cel W_j^- \asgn [m] \smin \Cel W_j^+$.
\itend
\item\label{step_Check1_Xi}
If for some $i_0\in \Cel W_j^-$ it holds that $f_{i_0}(x,y)=\top$, then do:
\itstart
\item $\Gamma_{j+1}\asgn \set{\gamma\in \Gamma_j}[i_0\in \gamma]$;
\item go to Step~\ref{step_LoopBeg_Xi}.
\itend
\item\label{step_Check2_Xi}
If for some $i_0\in \Cel W_j^+$ it holds that $f_{i_0}(x,y)=\bot$, then do:
\itstart
\item $\Gamma_{j+1}\asgn \set{\gamma\in \Gamma_j}[i_0\nin \gamma]$;
\item go to Step~\ref{step_LoopBeg_Xi}.
\itend
\item\label{step_Answ_Xi}
Output $W_j^+$ and \ebi{halt}.
\enend
\end{mdframed}

We claim that $\Xi(x,y)$ has complexity \asO{k(n)^6\tm\ell(n)} and outputs the set $F(x,y) = \set{i}[f_i(x,y) = \top]$.

\para[par_Xi_Gj]{At the end of Step~\ref{step_LoopBeg_Xi}
\m{
\set{i}[f_i(x,y) = \top] \in \Gamma_j
}
always.
}
In the beginning this is trivially true, and the updates (shrinkages) of $\Gamma$ in Steps~\ref{step_Check1_Xi} and \ref{step_Check2_Xi} occur under conditions that guarantee that $\set{i}[f_i(x,y) = \top]$ stays inside $\Gamma_{j+1}$.

\para{The answer is always correct.}
If the conditions of steps both~\ref{step_Check1_Xi} and \ref{step_Check2_Xi} where unsatisfied, then it must be the case that
\m{
W_j^+ = \set{i}[f_i(x,y) = \top]
}
in Step~\ref{step_Answ_Xi}.

\para{The protocol makes \asO{\ell(n)} iterations.}
It follows from the definitions of $W_j^+$ and $W_j^-$ that
\m{
\sz{\Gamma_{j+1}} \le \dr{\sz{\Gamma_j}}2
}
if the condition of either Step~\ref{step_Check1_Xi} or Step~\ref{step_Check2_Xi} is satisfied, \paref{par_Xi_Gj} guarantees that $\Gamma_j \ne \es$ and it is assumed by the theorem statement that $\sz{\Gamma_1} \le 2^{\ell(n)}$.

\para{The \Pxy-complexity of one iteration of the protocol is in \asO{k(n)^6}.}
To perform the check of Step~\ref{step_Check1_Xi} we use the protocol guaranteed by \crlref{crl_PNP_i} with respect to
\m{
f'(x,y) \deq \lor_{i\in W_j^-}^mf_i(x,y)
,}
and for the check in Step~\ref{step_Check2_Xi} we use it with
\m{
f''(x,y) \deq \lor_{i\in W_j^+}^m \lnot f_i(x,y)
,}
where $\lnot f_i(\dt,\dt)$ stands for the negation of the predicate $f_i(\dt,\dt)$.
Clearly, the functions $f',f'':\: \Cl A \times \Cl B \to \TF$ satisfy the requirements of \crlref{crl_PNP_i} and the corresponding $i_0$ meets the needs of Steps~\ref{step_Check1_Xi} and \ref{step_Check2_Xi}, respectively.
\prfend[\theoref{t_PNP_search}]

\sect[s_three]{Three-party communication with listening Charlie}

Let $g(x,y,z):\: \Cl A \times \Cl B \times \Cl C \to\OI$ be a tripartite total function.\fn
{
We are now using $\OI$ as the default range in the definitions of the communication problems (as opposed to the previously used $\TF$) as they no longer possess any ``logical asymmetry'':\ in the constructions of \sref{s_patt} that resulted from the asymmetry in the standard notion of computational non-determinism.
Moreover, the results of this part would remain valid if the considered problem were a \e{tripartite total function with any range} (the proof of \theoref{t_XYZ} would be based on the same idea but phrased somewhat differently if $\sz{\set{g(x,y,z)}_{x,y,z}}\in\asom1$).
}
The three input values will always be partitioned according to the \e{``number in hand''} input partition, that is, Alice receives $x$, Bob receives $y$ and Charlie receives $z$.

The only three-party communication regime that we will consider in this work is \e{deterministic}, therefore we will drop ``\Pxy'' and only depict the ``communication layout'' of each model in the corresponding notation.
For that we will deliberately use a not-too-abbreviated, but hopefully, rather intuitive layout representation (cf.~Footnote~\ref{fn_notation}).

\ndefi[d_XiYtZ]{\XiYtZ, interacting Alice and Bob with listening Charlie}{
For $n\in\NN$, let the sets $\sz{\Cl A}, \sz{\Cl B}, \sz{\Cl C}$ be such that $\Max{\sz{\Cl A}, \sz{\Cl B}, \sz{\Cl C}}\in (2^{n-1}, 2^n]$ and let $g:\: \Cl A \times \Cl B \times \Cl C \to \OI$.

Let $\Pi$ be a deterministic protocol where
\itstart
\item Alice receives $x$, Bob receives $y$ and Charlie receives $z$;
\item Alice and Bob interact in the broadcasting regime, that is, Charlie receives the transcript of their communication;
\item Charlie produces the answer.
\itend
If the transcript of $\Pi(x,y,z)$ contains at most $k(n)$ bits and the protocol computes $g(x,y,z)$, then we say that the \XiYtZ-complexity of $g$ is at most $k(n)$.
We say that $g$ is efficiently computable in \XiYtZ\ if its complexity in the model is at most $\plog(n)$.

Alternatively, the model \XiYtZ\ can be described as letting Alice and Bob interact in order to produce a message that is sent to Charlie, who must answer upon receiving it.
}

The following models could be pictured as ``merging'' a pair of players (or allowing free communication between them).
In those cases we will address the merged player by the name of the first individual (e.g., ``Alice+Bob'' -- the player who receives input $(x,y)$ -- will be called Alice and so on).

\ndefi[d_XYtZ]{\XYtZ}{
An \XYtZ-protocol is a deterministic protocol where
\itstart
\item Alice receive $(x,y)$; Charlie receives $z$;
\item Alice send a message to Charlie;
\item Charlie produces the answer.
\itend
If it computes $g:\: \Cl A \times \Cl B \times \Cl C \to\OI$, then we say that the \XYtZ-complexity of $g$ is at most the maximum number of bits sent by this protocol for the given input length.
The notion of efficiency for \XYtZ\ is similar to that for \XiYtZ\ (cf.~\defiref{d_XiYtZ}).
}

\ndefi[d_XZiYZ]{\XZiYZ}{
An \XZiYZ-protocol is a deterministic protocol where
\itstart
\item Alice receives $(x,z)$ and Bob receives $(y,z)$;
\item they interact;
\item Bob produces the answer.
\itend
If it computes $g:\: \Cl A \times \Cl B \times \Cl C \to\OI$, then we say that the \XZiYZ-complexity of $g$ is at most the maximum number of bits sent by this protocol for the given input length.
The notion of efficiency for \XZiYZ\ is similar to that for \XiYtZ\ (cf.~\defiref{d_XiYtZ}).
}

\ndefi[d_XYZcap]{\XYZcap}{
For $n\in\NN$, let the sets $\sz{\Cl A}, \sz{\Cl B}, \sz{\Cl C}$ be such that $\Max{\sz{\Cl A}, \sz{\Cl B}, \sz{\Cl C}}\in (2^{n-1}, 2^n]$ and let $g:\: \Cl A \times \Cl B \times \Cl C \to \OI$.

The \XYZcap-complexity of $g$ is the maximum of its \XZiYZ- and \XYtZ-complexities; if that is at most $\plog(n)$, then we say that $g$ is efficiently computable in \XYZcap. 
}

Recall that the models both \XZiYZ\ and \XYtZ\ can be viewed as \e{amplifications of \XiYtZ} obtained via letting a pair of players communicate for free; accordingly, solving a communication problem in \XiYtZ\ is at least as hard as solving it in \XYZcap.

\theo[t_XYZ]{
Let $g:\: \Cl A \times \Cl B \times \Cl C \to\OI$ be a tripartite total function whose \XZiYZ-complexity is $k(n)$ and \XYtZ-complexity is $\ell(n)$, then the \XiYtZ-complexity of $g$ is in \asO{k(n)^6\tm\ell(n)}.
In particular, $g(x,y,z)$ has an efficient \XiYtZ-protocol if and only if it is efficiently computable in \XYZcap.
}

\prfstart
As both \XZiYZ\ and \XYtZ\ are amplifications of \XiYtZ, the existence of an efficient protocol in the latter model trivially implies efficient computability in \XYZcap.

Consider an \XYtZ-protocol $\Pi_1$ and let $\alpha_{x,y}\in \OI^{\ell(n)}$ denote the message sent by Alice to Charlie when her input is $(x,y)\in \Cl A \times \Cl B$.\,\fn
{
Recall that we are addressing ``merged'' players by the name of the first included individual.
}
As receiving this message allows Charlie to compute $g(x,y,z)$, every possible message $\alpha_{x,y}$ corresponds to a function $\Cl C \to\OI$:\ this function is the description of Charlie's behaviour when he receives the corresponding message from Alice and $z\in \Cl C$ as input.

For every $z_0\in \Cl C$, let
\m{
f_{z_0}(x,y) \deq
\Cases
{\top}{if $g(x,y,z_0)=1$;}
{\bot}{otherwise.}
}
We will apply \theoref{t_PNP_search} to the family $\lf(f_z\rt)_{z\in \Cl C}$.
On the one hand, the \Pxy-complexity of $f_{z_0}(x,y)$ is the \XZiYZ-complexity of computing $g(x,y,z_0)$ when $(x,y)\in \Cl A \times \Cl B$, which is at most $k(n)$.
On the other hand, for every $(x_0,y_0)\in \Cl A \times \Cl B$ it holds that
\m{
\set{z\in \Cl C}[f_z(x_0,y_0)=\top]
= \set{z}[g(x_0,y_0,z)=1]
= \set{z}[\alpha_{x_0,y_0}(z)=1]
,}
where we let ``$\alpha_{x_0,y_0}(\dt)$'' stand for the function in $\Cl C \to\OI$ that the corresponding message represents, as discussed above.
Accordingly, every such set corresponds to some $\alpha_{x,y}\in \OI^{\ell(n)}$ and
\m{
\sz{
\set[:]{\set{z\in \Cl C}[f_z(x_0,y_0)=\top]}[(x_0,y_0)\in \Cl A \times \Cl B]
} \le 2^{\ell(n)}
.}

\theoref{t_PNP_search} guarantees the existence of a deterministic bipartite protocol of cost \asO{k(n)^6\tm\ell(n)} that receives $(x,y)\in \Cl A \times \Cl B$ and computes the set $\set{z_0\in \Cl C}[f_{z_0}(x,y)=\top]$.
In our \XiYtZ-protocol Alice and Bob will use that procedure, then send to Charlie some $\alpha_{x',y'}\in \OI^{\ell(n)}$ that corresponds to that set, that is,
\m{
\set{z_0\in \Cl C}[f_{z_0}(x,y)=\top]
= \set{z_0}[\alpha_{x',y'}(z_0)=1]
.}
Upon receiving it, Charlie, who knows $z$, will answer with $\alpha_{x',y'}(z) = g(x,y,z)$.
\prfend[\theoref{t_XYZ}]

\sect[s_conc]{Conclusions}

The study of communication complexity was initiated by Abelson~\cite{A78_Lo}, it was aimed at ``\e{assessing the complexity of computations carried out in distributed networks}''.
Since then our understanding of the area has somewhat advanced, so it may be desirable to summarise the achievement and to identify new interesting directions.
This work has been motivated both by the original question due to P.\ Hrube\v{s} and by the latter goal.

We saw a new structural result in the context of bipartite communication complexity of total functions:\ on the one hand, it could be viewed as a rather natural generalisation of what was known previously; on the other hand, it had somewhat non-trivial implications for the multi-party case.\fn
{
The three-party construction from \sref{s_three} can possibly be generalised for more participants.
}
The main theme of this work was looking for limitations that were imposed by the assumed \e{total structure} of the communication problem upon the ``structural diversity'' of communication complexity classes:
\itstart
\item in the two-party case we've seen that the newly defined class \PNPxy\ -- a generalisation of previously studied \UPxy\ and \FPxy\ -- admits efficient deterministic protocols, i.e., $\PNPxy \sbseq \Pxy$;
\item in the multi-party case we've seen that the class of tripartite total functions efficiently computable by deterministic interacting Alice and Bob with listening Charlie (the model that we denoted by \XiYtZ) equals the weakest among its two-party amplifications obtained by allowing free communication between a pair of players.
\itend
Previously known examples of such limitations are $\NPxy \cap \coNPxy \sbseq \Pxy$~\cite{AUY83_On}, $\UPxy \sbseq \Pxy$~\cite{Y91_Ex_Co} and $\FPxy \sbseq \Pxy$~\cite{KNSW94_Non}.
None of these five inclusions (in fact, equalities) among the classes would hold if the functions to be computed were not total.

From the combinatorial standpoint, the case of total functions is very natural in the context of communication complexity.
What other structural implications does it have?
In particular, what are the ``strengths of determinism'' that are exclusive for total functions?

\toct{Acknowledgements}

\sect*{Acknowledgements}
I am grateful to Pavel Hrube\v{s} both for the original motivating question and for numerous insightful discussions.
A number of very useful suggestions, both technical and editorial, have been received from colleagues and anonymous reviewers.

\toct{References}

\end{document}